\documentclass[fleqn,usenatbib]{mnras}

\usepackage{newtxtext,newtxmath}

\usepackage[T1]{fontenc}

\DeclareRobustCommand{\VAN}[3]{#2}
\let\VANthebibliography\thebibliography
\def\thebibliography{\DeclareRobustCommand{\VAN}[3]{##3}\VANthebibliography}

\usepackage{graphicx}	
\usepackage{amsmath}	
\usepackage{siunitx}    
\usepackage{subfig}     
\usepackage{newtxtext,newtxmath}
\usepackage{textgreek}  

\newcommand{\heii}{He\,\textsc{ii}}
\newcommand{\lya}{Ly\textalpha}
\newcommand{\nv}{N\,\textsc{v}}
\newcommand{\civ}{C\,\textsc{iv}}
\newcommand{\ciii}{C\,\textsc{iii}]}
\newcommand{\oii}{[O\,\textsc{ii}]}
\newcommand{\oiii}{[O\,\textsc{iii}]}
\newcommand{\oiiiuv}{O\,\textsc{iii}]}
\newcommand{\niii}{N\,\textsc{iii}]}

\newcommand{\flux}{erg\,s$^{-1}$\,cm$^{-2}$}
\newcommand{\kms}{\SI{}{\km\per\second}}
\newcommand{\fesc}{$f_\textrm{esc}$}

\newcommand{\xiion}{$\xi_{\textrm{ion}}$}
\newcommand{\ew}{EW$_0$}

\title[C\,\textsc{iv} emission at $z\sim3.5$]{Strong C\,\textsc{iv} emission from star-forming galaxies: a case for high Lyman continuum photon escape}

\author[A. Saxena et al.]{A. Saxena,$^{1}$\thanks{E-mail: aayush.saxena@ucl.ac.uk}
E. Cryer,$^{1}$
R. S. Ellis,$^{1}$
L. Pentericci,$^{2}$
A. Calabr\`o,$^{2}$
S. Mascia,$^{2}$
A. Saldana-Lopez,$^{3}$
\newauthor D. Schaerer,$^{3,4}$
H. Katz,$^{5}$
M. Llerena$^{6}$
and R. Amor\'{i}n$^{6,7}$
\\
$^{1}$Department of Physics and Astronomy, University College London, Gower Street, London WC1E 6BT, UK\\
$^{2}$INAF – Osservatorio Astronomico di Roma, via Frascati 33, 00078, Monteporzio Catone, Italy\\
$^3$Department of Astronomy, University of Geneva, 51 Chemin Pegasi, 1290 Versoix, Switzerland\\
$^4$CNRS, IRAP, 14 avenue E. Belin, F-31400 Toulouse, France\\
$^5$Astrophysics, University of Oxford, Denys Wilkinson Building, Keble Road, Oxford OX1 3RH, UK\\
$^6$Departamento de F\'isica y Astronomi\'a, Universidad de La Serena, Av. Juan Cisternas 1200 Norte, La Serena, Chile\\
$^7$Instituto de Investigaci\'on Multidisciplinar en Ciencia y Tecnolog\'ia, Universidad de La Serena, Ra\'ul Bitr\'an, 1305 La Serena, Chile
}

\date{Accepted XXX. Received YYY; in original form ZZZ}

\pubyear{2022}

\begin{document}
\label{firstpage}
\pagerange{\pageref{firstpage}--\pageref{lastpage}}
\maketitle

\begin{abstract}
Finding reliable indicators of Lyman continuum (LyC) photon leakage from galaxies is essential in order to infer their escape fraction in the epoch of reionisation, where direct measurements of LyC flux are impossible. To this end, here we investigate whether strong C\,\textsc{iv}\,$\lambda \lambda 1548,1550$ emission in the rest-frame UV spectra of galaxies traces conditions ripe for ample production and escape of LyC photons. We compile a sample of 19 star-forming galaxies in the redshift range $z=3.1-4.6$ from the VANDELS survey that exhibit strong C\,\textsc{iv} emission, producing a stacked spectrum where all major rest-UV emission lines are clearly detected. Best-fitting spectral energy distribution models containing both stellar and nebular emission suggest the need for low stellar metallicities ($Z=0.1-0.2\,Z_\odot$), young stellar ages ($\log(\rm{age/yr}) = 6.1-6.5$), a high ionisation parameter ($\log U = -2$) and little to no dust attenuation ($E(B-V)=0.00-0.01$). However, these models are unable to fully reproduce the observed C\,\textsc{iv} and He\,\textsc{ii} line strengths. We find that the Ly$\alpha$ line in the stacked spectrum is strong and peaks close to the systemic velocity, features that are indicative of significant LyC photon leakage along the line-of-sight. The covering fractions of low-ionisation interstellar absorption lines are also low, implying LyC escape fraction in the range $\approx 0.05-0.30$, with signatures of outflowing gas. Finally, C\,\textsc{iv}/C\,\textsc{iii}] ratios of $>0.75$ for a subset of individual galaxies with reliable detections of both lines are also consistent with physical conditions that enable significant LyC leakage. Overall, we report that multiple spectroscopic indicators of LyC leakage are present in the stacked spectrum of strong C\,\textsc{iv} emitting galaxies, potentially making C\,\textsc{iv} an important tracer of LyC photon escape at $z>6$.
\end{abstract}




\section{Introduction}
\label{sec:intro}
Understand the contribution of star-forming galaxies in governing cosmic reionisation, a process whereby the intergalactic medium (IGM) underwent a phase transition from a neutral to completely ionised gas, requires measures of their ionising photon production efficiencies (\xiion) the fraction of Lyman continuum (LyC; $\lambda_0 < 912$\,\AA) photons that manage to escape from the galaxy (\fesc; see \citealt{day18} for a review) and the integrated space density of galaxies, derived from UV luminosity functions \citep[e.g.][]{rob13, rob15, bou15, fin15}.

The nature and strength of the ionising radiation emerging from young stars within star-forming galaxies can be derived from emission lines seen at rest-frame UV and optical wavelengths \citep[e.g.][]{gut16, fel16, xia18, pla19}, including the Balmer H$\alpha$ line whose intensity is related to the number of ionising photons produced through recombination physics \citep[e.g.][]{shi18}. Detailed spectroscopic studies of galaxies at intermediate redshifts have provided reliable measurements of \xiion\ and we can expect continued progress within the reionisation era ($z\gtrsim6$) following the successful deployment of the \emph{James Webb Space Telescope (JWST)}. 

Due to the increasing neutrality of the IGM at higher redshifts, however, direct measurements of LyC radiation from galaxies become challenging at $z\gtrsim4$ \citep[e.g.][]{ino14}. A popular strategy for estimating \fesc\ for galaxies in the reionisation era is to study `analogues' (i.e. with comparable physical properties) of $z>6$ galaxies  at intermediate redshifts ($z\sim3$), for which LyC leakage can be directly observed \citep[e.g.][]{sha06}. With a detailed understanding of the ionising output of stars and conditions of the interstellar medium (ISM) that enable significant escape of LyC photons \citep[e.g.][]{nak20, sax22} in analogous intermediate redshift galaxies, it may then be possible to infer which types of sources are likely to be the key drivers of cosmic reionisation.

Although several dedicated searches have identified LyC leaking galaxies from large area surveys at intermediate redshifts, their number remains quite modest \citep[see][for a recent compilation]{mes21}. Inference from ground-based data, in particular, is restricted mostly to luminous galaxies \citep[e.g.][]{gra16, gua16, mar17, nai18, ste18, sax22}. A further complication in assessing the statistics of LyC leakers in any given survey is the expectation from numerical simulations that LyC leakage may be highly anisotropic, such that detections are only possible when the leaking channels are favourably aligned to the line-of-sight to the observer \citep[e.g.][]{kat20, bar20, kim22}, with some observational evidence to support this idea \citep[e.g.][]{van21}.

Rather than undertaking a systematic survey for LyC leakage in a sample of photometrically-selected star forming galaxies, it may be more productive to study galaxies whose emission line properties are indicative of emission of copious amounts of ionising photons with high ionisation parameters. Specific examples include sources with high \oiii\,$\lambda5007$/\oii\,$\lambda3727$ ratios \citep[e.g.][]{nak18, izo18a}, which may reflect density-bound nebulae from which LyC leakage may be possible \citep{zac13, nak14}, as well as those with high ionisation energy lines which may reflect young stellar populations whose associated supernovae can clear channels that permit free passage of LyC photons \citep{ber19, nan19, sax20, tan21, van21, sen21}.

Following this strategy, in this paper we focus on consideration of the \civ\,$\lambda\lambda\, 1548,1550$ doublet, whose ionisation energy is $47.9$\,eV and has already been detected in rest-UV spectra of several $z>6$ galaxies (e.g. \citealt{star15, mai17, sch17}, although it remains unclear whether the \civ\ emission is purely due to star-formation or due to active galactic nuclei or AGN). In the local universe, strong nebular \civ\ emission is almost exclusively seen in low-mass galaxies with extremely low metallicities ($\lesssim 0.1\,Z_\odot$, where $Z_\odot$ is the solar metallicity value), young stellar ages ($\log(\rm{age/yr}) \lesssim 7$) and high specific star-formation rates ($\log(\rm{sSFR/yr}^{-1}) < -8$; \citealt{ber16, ber18, sen17, sen19}), properties that are likely common amongst galaxies in the reionisation era. 

There is also growing evidence that strong \civ\ emission may be associated with LyC leakage: \citet{sch22} found strong \civ\ emission ubiquitously in a sample of $z<0.7$ galaxies, and strong \civ\ has also been observed in a confirmed LyC leaker at $z\sim3$ \citep[e.g.][]{van16a}. \citet{sch22} interpreted their results with an elevated level of \xiion\ and the presence of density-bound H\,\textsc{ii} regions. At low gas-phase metallicities, these conditions increase both the \civ\ luminosity as well as the \civ/\ciii\ line ratio. Additionally, as a resonant line, \civ\ may trace photon escape through high-ionisation gas \citep{ber19} and its typical P-Cygni profile is a valuable indicator of outflows driven by massive stars \citep[e.g.][]{ste16} that are necessary to clear out channels for LyC escape. All the foregoing suggests \civ\ emission may be an important pointer to LyC leakage.

In this paper we investigate the properties of a sample of star-forming galaxies with spectroscopic redshifts at $z\sim3.1-4.6$ selected purposely to have strong \civ\ emission. Our aim is to better understand the spectroscopic properties of strong \civ\ emitting galaxies and via independent spectroscopic measures, investigate the presence of signatures that may point towards significant LyC photon leakage. Our study complements ongoing efforts to understand the properties of star-forming galaxies that leak LyC photons at lower redshifts (e.g. Low-Redshift Lyman Continuum Survey, \citealt{flu22a, flu22b} and the COS Legacy Archive Spectroscopy Survey, \citealt{ber22}). Ultimately, our study aims to provide a reference for an improved understanding and interpretation of \emph{JWST}/NIRSpec spectra of star-forming galaxies at $z>6$ in the context of LyC photon escape.

The layout of this paper is as follows. We describe the spectroscopic data, identification of \civ\ emitting galaxies, emission line measurements and flagging of potential AGN in \S\ref{sec:data}. We present a stacked spectrum of \civ\ emitting galaxies along with a detailed analysis of other strong rest-frame UV emission lines in \S\ref{sec:stack}. We explore the nature of the underlying sources of ionisation that best describe the observed \civ\ emission (and other lines) in the stack in \S\ref{sec:ionisation}. Finally, we investigate whether significant LyC photon leakage can be inferred from the stacked spectrum of \civ\ emitting galaxies using other indirect spectral signatures in \S\ref{sec:fesc}, summarising our findings in \S\ref{sec:summary}.

Throughout this paper, we assume a $\Lambda$CDM cosmology with $\Omega_m = 0.3$ and $H_0 = 67.7$ \kms\,Mpc$^{-1}$ taken from \citet{planck}. All logarithms are in base 10, unless otherwise specified. In this paper we adopt a solar metallicity value of $Z_\odot = 0.02$.

\section{Data}
\label{sec:data}
We use spectroscopic data from VANDELS -- a deep VIMOS survey of the CANDELS fields -- which is a recently completed ESO public spectroscopic survey carried out using the VLT. VANDELS covers two well-studied extragalactic fields, the UKIDSS Ultra Deep Survey (UDS) and the Chandra Deep Field South (CDFS/GOODS-S). We refer the readers to \citet{mcl18} for details about the survey description and target selection, and to \citet{pen18} for more information about data reduction and spectroscopic redshift determination. The final VANDELS data release, DR4\footnote{\url{http://vandels.inaf.it/dr4.html}}, contains spectra of $\sim2100$ galaxies in the redshift range $1.0<z<7.0$, with on-source integration times ranging from 20 to 80 hours, where $>70\%$ of the targets have at least 40 hours of integration time \citep{gar21}. The spectral resolution of VANDELS spectra is $R\sim600$.

The reliability of redshifts in the VANDELS database is recorded using the following flags: 0 -- no redshift could be assigned, 1 -- 50\% probability to be correct, 2 -- 70-80\% probability to be correct, 3 -- 95-100\% probability to be correct, 4 -- 100\% probability to be correct and 9 -- spectrum shows a single emission line. The typical accuracy of spectroscopic redshift measurements is $\sim150$ km~s$^{-1}$ \citep{pen18}.

\subsection{A search for \civ\ emitters}
In this work we only select spectroscopically confirmed galaxies from VANDELS that have a redshift reliability flag of either 3 or 4, which guarantees that the redshift measured by the VANDELS team has a $>95$\% probability of being correct. This also ensures reliable detection of other emission or absorption features in the spectrum. 

Since our main goal is to explore spectroscopic properties of galaxies that show strong \civ\ $\lambda\lambda 1548,1550$, we limit our focus to the redshift range $z=3.1-4.6$ where additionally the \lya\ line is visible. We find 735 galaxies across the CDFS and UDS fields in the redshift range $z=3.1-4.6$ and redshift reliability flags 3 or 4, which constitute the parent sample in this study.

\civ\ emission in this work is identified primarily via visual inspection. We first inspect the 1D spectra in the parent sample to search for \civ\ emission. If the \civ\ line coincides with a skyline residual, we discard the source as any line flux measurement would be relatively unreliable. Once \civ\ emission is identified in the 1D spectrum, we inspect the 2D spectrum to ensure that the emission is real and not due to hot pixels, noise peaks and/or sky residuals. Two of the co-authors independently performed the visual identification and only those that were identified as \civ\ emitters by both individuals from 1D and 2D spectra were retained in our sample for further analysis. Out of 735 galaxies in the parent sample, we identified 22 sources as reliable \civ\ emitters. None of these are detected in the deep \emph{Chandra} X-ray catalogues in either CDFS \citep{luo17} or in UDS \citep{koc18} fields, ruling out any clear AGN activity.

We note here that the goal of this study is not to identify a complete sample of \civ\ emitters in the VANDELS survey, but to explore the spectroscopic properties of sources that show clear evidence of strong \civ\ emission at intermediate redshifts. Therefore, we do not include galaxies with marginal \civ\ detections, or possible line emission that may be partly contaminated by a sky residual and would otherwise be considered as a real \civ\ line. A more statistically complete sample of \civ\ emitters from the VANDELS survey will be assembled in a future study (Mascia et al. in prep).

\subsection{Rest-UV emission line measurements}
\label{sec:lines}
We measure line fluxes of \civ\ and other rest-frame UV emission lines visible in the spectra, which include \lya, \heii\ $\lambda1640$ and \oiiiuv\ $\lambda\lambda 1660, 1666$. The line fitting is performed using the \textsc{python} package \textsc{mpdaf}\footnote{\url{https://mpdaf.readthedocs.io/en/latest/index.html}} in a similar fashion to \citet{sax20}. Briefly, we fit the observed emission lines with single or double Gaussian functions and measure the local continuum level in a wavelength range free of other emission or absorption either side of the line, calculating the line fluxes, full width at half maxima (FWHM) and rest-frame equivalent widths (EW$_0$). Below we give a summary of the rest-frame UV lines identified in the individual spectra in this work.

By design, we identify \civ\ emission in all 22 objects. The integrated line fluxes range from $0.2 - 6.0\times10^{-18}$\,\flux, with rest-frame equivalent widths ranging from EW$_0$ $=0.7-12.8$\,\AA.

18 out of 22 objects show clear \lya\ emission with a range of line strengths, widths and profiles. Three sources do not show any \lya\ emission in their spectra and the \lya\ line in one source is contaminated by a sky line.

Nine sources also show \heii\ emission with signal-to-noise ratio (S/N) $\geq2$, five of which were also identified by \citet{sax20} using an earlier data release of VANDELS. The integrated \heii\ line fluxes range from $0.2 - 7.4 \times 10^{-18}$\,\flux\ and EW$_0$ in the range $2.0 - 18.7$\,\AA. 

Nine sources show the \oiiiuv\ $\lambda\lambda1660,1666$ emission lines as well, which often appear to be blended and therefore only the total \oiiiuv\ flux is measured and reported. The \oiiiuv\ line fluxes range from $0.4-7.2 \times 10^{-18}$\,\flux, with EW$_0$ ranging from $1.5-11.3$\,\AA.

The wavelength coverage of VANDELS spectra allows the detection of \ciii\,$\lambda \lambda 1907, 1909$ (which appear to be blended) only at $z\lesssim3.9$ in our sample. Additionally, since the \ciii\ line lies in a relatively red part of the observed spectrum, it is more prone to contamination by skyline residuals. Therefore, we only robustly identify \ciii\ in 5 objects, with line fluxes ranging from $0.7-4.3 \times 10^{-18}$\,\flux, and EW$_0$ in the range $3.3-17.2$\,\AA, which is comparable to a more statistical measurement of \ciii\ from VANDELS galaxies presented by \citet{lle21}.

The full suite of \lya, \civ, \heii, and \oiiiuv\ lines are detected with S/N $\geq2$ for only three sources in our sample, making it necessary to perform spectral stacking to boost S/N of other rest-UV features necessary to characterise the properties of the underlying sources of ionising photons (see \S\ref{sec:stack}). However, before producing a stacked spectrum of \civ\ emitters, we attempt to identify sources that may be dominated by AGN activity based on their \civ\ and \heii\ line strengths.

\subsection{Identifying possible AGN}
\label{sec:agn}
AGNs are capable of producing a much larger number of ionising photons through the accretion of material on to the central supermassive black hole, which results in a non-thermal spectral energy distribution that can easily produce extremely high energy photons giving rise to higher order transitions of the most common elements, such as \civ, \heii\ and \nv.

Since our sample is selected on the \civ\ line and the second most common nebular emission line seen in individual galaxy spectra is \heii, we use model predictions built around these two lines to identify possible AGN in our sample. In particular, we employ the diagnostic from \citet{nak18}, comparing the \civ/\heii\ ratio against \ew(\civ). The \citet{nak18} emission line predictions are obtained using the photoionisation code Cloudy \citep{fer13}. To model the spectral energy distribution (SED) of star-forming galaxies, the authors employ BPASS models \citep{stan15} that include the effect of interacting binary stars. The AGN models on the other hand assume a power law ionising radiation field emanating from the narrow-line region surrounding the active black hole.

\begin{figure}
    \centering
    \includegraphics[width=\linewidth]{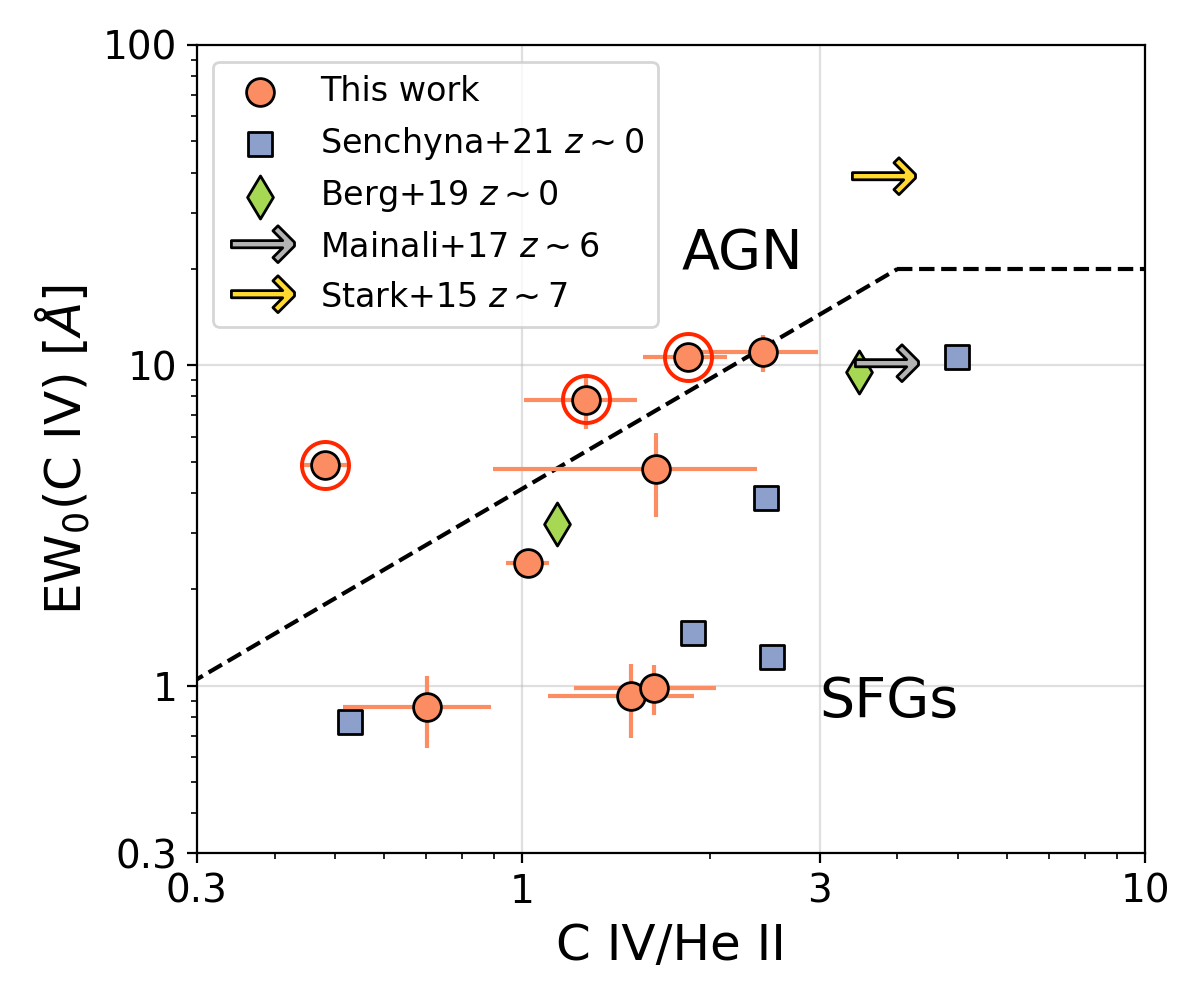}
    \caption{Distribution of \civ/\heii\ flux ratio versus \civ\ rest-frame equivalent width for individual galaxies in this work where both emission lines are securely detected. The division between ionisation due to star-formation alone (SFG) and ionisation due to AGN has been adapted from the modelling of \citet{nak18}. Also shown for comparison are measurements from local low mass, low metallicity galaxies \citep{ber19, sen21}, as well as the \civ\ detection from a galaxy at $z\sim7$ reported by \citet{star15} and at $z\sim6$ reported by \citet{mai17}. We find that three \civ\ emitters are likely to be dominated by AGN that are marked with red circles, and six galaxies can be explained through star-formation activity alone, lying close to the AGN-SFG dividing line implying the presence of sources hard ionising radiation. The three sources identified as AGN are removed from our final sample.}
    \label{fig:agn-sfg}
\end{figure}

In Figure \ref{fig:agn-sfg} we show \ew(\civ) and \civ/\heii\ measured in the spectra of galaxies in our sample where both lines were detected with S/N\,$>2$. The dashed line demarcates the parameter space permitted by ionisation due to star-formation alone and ionisation due to AGN. For comparison, we also show measurements from low metallicity dwarf galaxies in the local Universe \citep{ber19, sen21}, as well as \civ\ measurements from a bright galaxy at $z\sim7$ reported by \citet{star15} and at $z\sim6.1$ reported by \citet{mai17}.

We identify three sources in our sample that are likely to be dominated by AGN, whereas six sources occupy the parameter space where star-formation alone is enough to explain the \civ\ and \heii\ emission. For sources with no \heii\ detection, the resulting lower limit on \civ/\heii\ $\gtrsim 2.5$ and \ew(\civ) $<10$\,\AA\ is suggestive of ionisation due to star-formation. Since our goal is to investigate \civ\ emission exclusively from star-forming galaxies, we remove these three potential AGN from our sample, leaving us with a final sample of 19 star-forming galaxies that show \civ\ emission in their spectra. Further, we do not find any individual detections for these sources in the deep X-ray data available in the fields.

We do note, however, that a few star-forming galaxies in our sample lie very close to the dividing line between photoionisation by AGN and star-formation. The parameter space occupied by our \civ\ and \heii\ emitting galaxies is also comparable to the low mass, metal-poor galaxies in the nearby Universe that show high ionisation lines. We also note that the very strong \civ\ detection in the $z\sim7$ galaxy reported by \citet{star15} suggests photoionisation due to AGN, however \civ\ measurement from the galaxy at $z\sim6.1$ from \citet{mai17} is comparable to what we find in our sample.

\begin{figure}
    \centering
    \includegraphics[width=\linewidth]{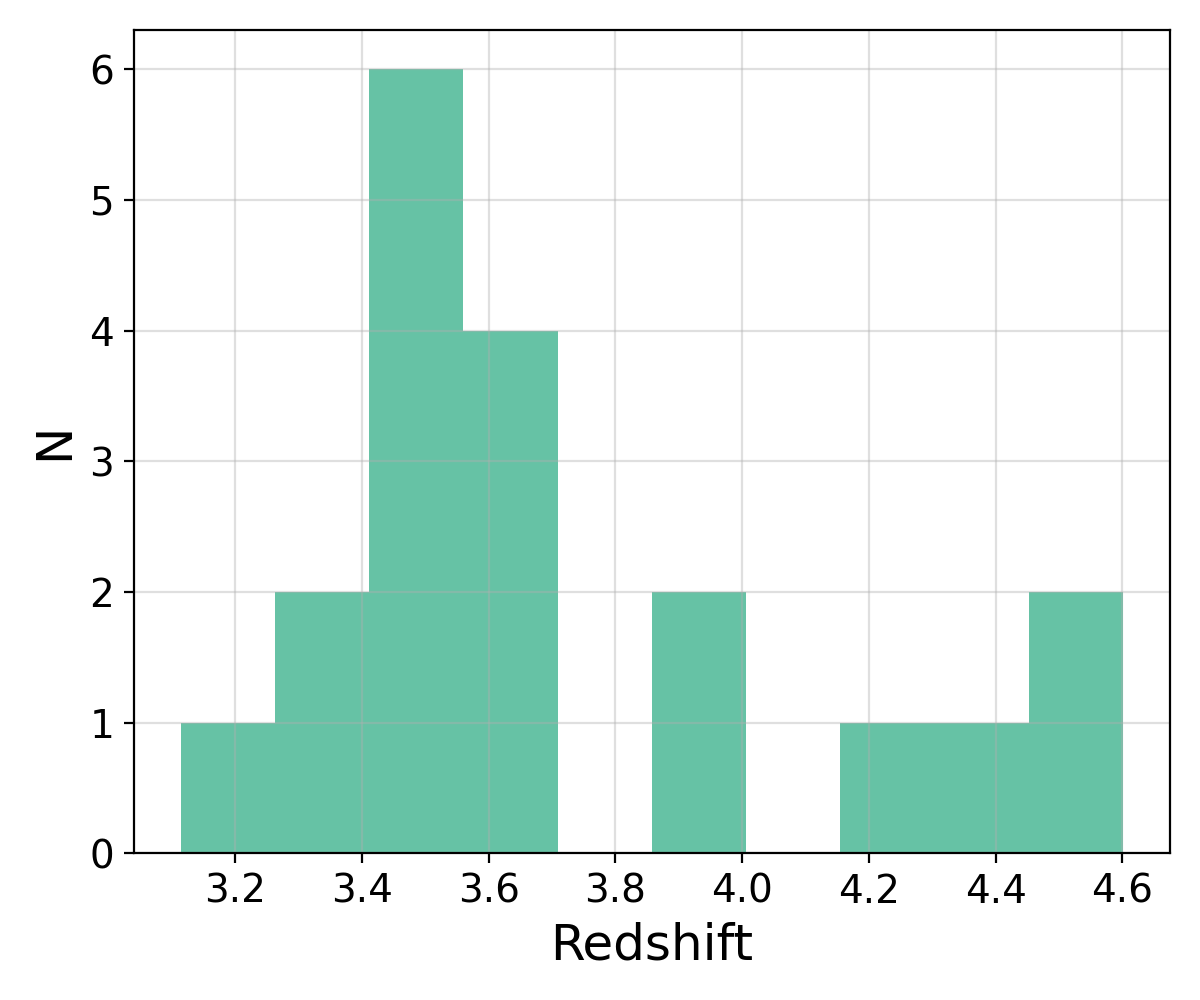}
    \caption{Redshift distribution of 19 \civ\ emitting star-forming galaxies in our final sample, selected from the VANDELS survey. The galaxies cover a redshift range of $3.1<z<4.6$, with a median redshift of $z=3.6$.}
    \label{fig:redshift}
\end{figure}

A histogram of the redshift distribution of 19 star-forming galaxies with strong \civ\ emission in our final sample is shown in Figure \ref{fig:redshift}. The median redshift of our sample is $3.6$ and the sources span an observed $i$-band magnitude range of $24.4-26.7$\,AB. Using these 19 galaxies, we now produce a stacked spectrum in the following section.

\section{A stacked spectrum of \civ\ emitters at $\mathbf{3<z<4.5}$}
\label{sec:stack}
As mentioned earlier, detecting the full suite of rest-UV emission and absorption features requires high S/N of the underlying continuum. Therefore, to boost S/N of these features to study the sample-averaged properties of \civ\ emitting galaxies and better understand their underlying stellar populations and state of the ISM, in this section we produce a stacked spectrum of \civ\ emitting galaxies.

\subsection{Stacking procedure}
The method to co-add spectra adopted in this study is similar to \citet{sax20} -- the stacking is performed by first de-redshifting each spectrum using the `systemic' redshifts derived primarily from the strong \civ\ emission lines, as well as \heii, \oiiiuv\ and \ciii\ lines whenever visible in the individual galaxy spectrum. For galaxies where two or more of the above-mentioned lines were detected, we found that the redshift derived from each line was within $100$\,\kms\ of each other, well within the resolution element of the spectrograph. The good agreement between the redshifts measured from these lines suggests the \civ\ line in emission generally traces systemic redshifts well across our galaxies, however this may not be universally true as the \civ\ emission line is prone to resonant scattering. 

The rest-frame spectra are normalised using the mean flux density value in the wavelength range $1460-1540$\,\AA. Each spectrum is then assigned a weight based on the standard deviation of flux in this wavelength range, where the weight is inversely proportional to the measured variance on the flux density at $\sim1500$\,\AA. Such an approach to calculating weights for stacking helps alleviate uncertainties that may be introduced by weighting the stacking using pixel-by-pixel standard deviations, which could be highly variable across individual sources. Assigning a weight based on the SNR at $1500$\,\AA, which is almost identical to luminosity weighted stacking, gives comparable results to that obtained by weighting using the standard deviation of pixels \citep[e.g.][]{cal22}.

The spectra are then re-sampled to a uniform wavelength grid ranging from 1050 to 1820\,\AA, which is the rest-frame wavelength range most commonly probed by observed spectra in our sample, with a step size of $0.56$\,\AA, which is the wavelength resolution obtained at a redshift of $3.6$, the median redshift of our final sample.

The wavelengths in the observed individual spectra that do not fall within the rest-frame wavelength grid for the stacked spectrum are masked on a source by source basis to avoid incomplete sampling of the fluxes across galaxies. This means that the \ciii\ line is unfortunately not covered in the stacked spectrum. We also mask residual sky lines and hot pixels in the spectra by employing a $>20\sigma$ clipping. A stacked spectrum is then produced using a weighted averaging procedure.

The standard deviation of the stacked spectrum will have contributions both from S/N limitations of individual spectra as well as source-by-source spectral variation in the individual sources \citep[e.g.][]{jon12}. Therefore, to capture both these effects, we use a bootstrapping method to calculate uncertainties on the stacked spectrum. To do this, we take our sample of 19 \civ\ emitters and replace one randomly selected \civ\ emitter by a \civ\ non-emitter selected from the parent VANDELS sample. We then produce a stacked spectrum of these 19 objects, with 18 \civ\ emitters and one non-emitter. We repeat this process of random replacement 500 times, consequently producing 500 stacked spectra with 18 \civ\ emitters and one non-emitter. 

\begin{figure*}
    \centering
    \includegraphics[width=\textwidth]{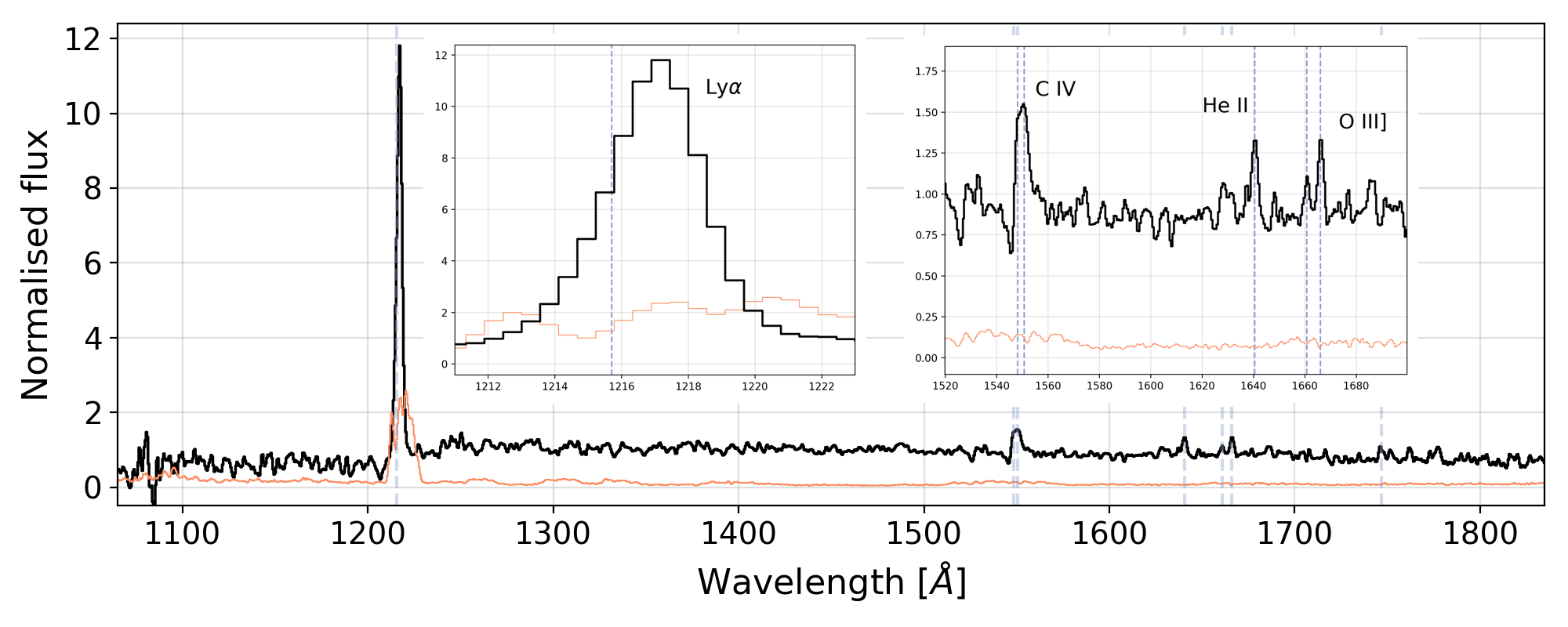}
    \caption{Stacked spectrum (black) of 19 \civ\ emitting star-forming galaxies in our final sample, with $1\sigma$ uncertainties (orange) obtained from bootstrapping. Multiple rest-frame UV emission lines are clearly detected in the stacked spectrum with high S/N. We show insets with the \lya\ line (left), and \civ\,$\lambda\lambda1548, 1550$, \heii\,$\lambda1640$ and \oiiiuv\,$\lambda\lambda1660,1666$ lines (right). We additionally mark the position of the blended \niii\ feature at $\sim \lambda1747$ line in the spectrum, which is not as strong as the other lines highlighted. The equivalent widths and FWHMs of emission lines in the stacked spectrum are given in Table \ref{tab:stack_lines}.}
    \label{fig:stack}
\end{figure*}

In this way, any strong spectral feature that is being contributed by an outlying \civ\ emitter can be accounted for, and the resulting standard deviation will be an accurate measure of the `true' deviation of the stacked spectrum. The standard deviation of these bootstrapped spectra is used to calculate the $1\sigma$ uncertainty on the final stacked spectrum of \civ\ emitters. The final stacked spectrum along with $1\sigma$ errors is shown in Figure \ref{fig:stack}.

A range of emission and absorption features are clearly detected in the stacked spectrum. In the sections that follow, we investigate the spectroscopic properties inferred from this stacked spectrum, and below we briefly summarise the observed properties of some of the brightest emission lines seen in the stack.

\subsection{Emission lines}
\label{sec:stack-lines}
The emission lines in the stack are measured in a manner similar to that discussed in \S\ref{sec:lines}. The errors on both the line flux and width measurements are a consequence of the $1\sigma$ uncertainty on the stacked spectrum calculated via bootstrapping.
 
The \lya\ line in the stack is strong and has a symmetric profile, with non-zero flux bluewards of the \lya\ peak. The peak of the \lya\ line is only slightly redshifted compared to the systemic redshift of the stacked spectrum. Such a line profile is typically associated with LyC emitting galaxies across redshifts \citep[e.g][]{ver17, gaz20}, which we investigate further in \S\ref{sec:lya}. A single Gaussian function provides a good fit to the line, with a measured FWHM of $820.0 \pm 17.0$\,km\,s$^{-1}$, and a high EW$_0 = 37.5 \pm 0.9$\,\AA, placing it in the regime of strong \lya\ emitting galaxies (LAEs). 

The \civ\ emission in the stacked spectrum is clear. Due to the relatively low spectral resolution the two components at $\lambda1548$ and $\lambda1550$ appear to be marginally blended, requiring a double Gaussian function fit to recover the full line flux. From this fit we recover a total \ew(\civ) $=5.2 \pm 0.5$. A P-Cygni absorption feature immediately bluewards of the $\lambda1548$ peak is also visible.

We identify strong \heii\ emission in the stacked spectrum and the emission line is fitted using a single Gaussian function, giving a FWHM(\heii) $=591.6 \pm 22.6$\,\kms\ and \ew(\heii) $=2.2 \pm 0.1$\,\AA. With He$^+$ ionising potential being $\approx54.4$\,eV, \heii\ emission is indicative of the presence of young stellar populations with low metallicities, capable of producing highly energetic photons.

We identify both \oiiiuv\,$\lambda1660,1666$ emission lines, with $\lambda1660$ being fainter than $\lambda1666$. The doublet is sufficiently separated in wavelength such that two independent Gaussian functions can be fit to the lines. We measure a FWHM of $518.3 \pm 25.9$\,\kms\ for $\lambda 1660$ and $611.0 \pm 30.8$\,\kms\ for $\lambda1666$. The total \ew(\oiiiuv) is $3.2 \pm 0.3$\,\AA, with a \oiiiuv\,$\lambda1660$/\oiiiuv\,$\lambda1666$ ratio of $\approx 0.5$.

We also find \niii\ emission around $1748$\,\AA, which is a quintuplet with lines at $1746, 1748, 1749, 1750$ and $1752$\,\AA, but appears to be blended in the spectrum. The \niii\ emission has a lower ionising potential of $\approx26.9$\,eV compared to \oiiiuv\ and other stronger lines visible in the stacked spectrum. The presence of \niii\ emission has not been widely reported -- including from stacked spectra -- in star-forming galaxies in the literature \citep[but see][]{lef19}. However, the presence of \niii\ is not surprising considering that the sources were selected on emission lines requiring much higher ionising energies. 

The \ew\ and FWHM of the above-mentioned lines are summarised in Table \ref{tab:stack_lines}. In the following section we compare line measurements from our stacks to other stacked spectra of star-forming galaxies produced at comparable redshifts, as well as \civ\ emitting galaxies identified both in the local Universe as well as at $z>6$. 
\begin{table}
    \centering
    \caption{Properties of emission lines clearly detected in the stacked spectrum.}

    \begin{tabular}{l c r}
    \hline 
    Line/Vacuum $\lambda$ (\AA) & EW$_0$ (\AA) & FWHM (\kms) \\
    \hline
    \lya\,$\lambda1216$ & $37.5 \pm 0.7$ & $830 \pm 18$  \\
    \civ\,$\lambda \lambda 1548, 1550$ & $5.2 \pm 0.5$ & $-$ \\
    \heii\,$\lambda1640$ & $2.2 \pm 0.1$ & $592 \pm 23$ \\
    \oiiiuv\,$\lambda1660$ & $1.1 \pm 0.3$ & $547 \pm 34$ \\
    \oiiiuv\,$\lambda1666$ & $2.0 \pm 0.1$ & $502 \pm 31$ \\
    \niii\,$\lambda1747$ & $1.2 \pm 0.3$ & $440 \pm 120$ \\
    
    \hline   \\
    
    \end{tabular}
    
    \textit{Note.} Since the \civ\ doublet appears to be blended and is fitted using a combination of two Gaussian functions, we do not report the FWHM.
    \label{tab:stack_lines}
\end{table}

\subsection{Comparison with literature}
\label{sec:literature}
The \ew(\lya) from our stack is comparable to that measured in the stacked spectrum of LBGs at $z\sim3$ from \citet{ste18} that were classified as LAEs (EW$_0$(\lya)\,$= 44.1$\,\AA), and the stack of the quartile of their sources with the strongest \lya\ emission, dubbed WQ4 (EW$_0$ (\lya)\,$= 43.2$\,\AA). Both LAEs and WQ4 galaxies in \citet{ste18} were identified to be the strongest LyC leakers using spectroscopic measurements. Interestingly, the strength of the \civ\ emission also appears to increase with increasing \lya\ in the stacked spectra of \citet{ste18}, but no measure of \civ\ emission across stacks is given.

Comparing with the stacks of narrow-band selected LAEs reported by \citet{nak18b}, we find that our \ew(\lya) is closest to that measured from the stack of UV-luminous LAEs having \ew(\lya)$= 38-40$\,\AA. Prominent \civ\ features were also reported in their stacks of LAEs with large \ew(\lya) and fainter UV magnitudes, with \ew(\civ) in the range $2.9-3.9$\,\AA. The \ew(\civ) measured by \citet{nak18b} even for their strongest LAEs is lower than what we observe.

We then compare our measurements with the stacked spectra of LAEs at $3<z<4.6$ in the MUSE HUDF reported by \citet{fel20}. The \civ\ doublet in emission is seen in their stack of LAEs that have low FWHM(\lya), high \ew(\lya) and faint UV magnitudes. The \ew(\civ) ranged from $1.95-4.74$\,\AA. A slightly weaker \heii\ emission with \ew(\heii) $\approx1-1.3$\,\AA\ was also reported from their stacks.

Only a handful of \civ\ observations currently exist at $z>6$. The \civ\ emitter identified by \citet{star15} at $z\sim7$ shows a higher \ew(\lya) $=65\pm12$\,\AA\ compared to our stack, a significantly higher \ew(\civ) $\approx40$\,\AA\ with \ew(\heii) $<11.4$\,\AA. However, there are suggestions that this source may be powered by an AGN, as we demonstrated in \S\ref{sec:agn}. The lensed \civ\ emitter at $z=6.11$ from \citet{sch17} also has a higher \ew(\lya) $=68 \pm 6$\,\AA\ and higher \ew(\civ) $=24 \pm 4$\,\AA\ compared to our stack. Another lensed \civ\ emitter at $z\sim6.1$ reported by \citet{mai17} has a comparable \ew(\lya) $=40\pm5$\,\AA\ with \ew(\civ) $\approx10$, which is higher than what is seen in our stack.

\civ\ emission was identified in an ultra-deep ($\approx 140$\,hr) MUSE spectrum of a galaxy at $z=4.77$ \citep{mat22b} with \ew(\civ)$ \approx 5$\,\AA\. \civ\ emission has also been found in a lensed galaxy at $z=4.88$ \citep{wit21} with \ew(\civ) $\approx18$\,\AA. These galaxies also show strong \lya\ emission with \ew(\lya) $\approx 62$\,\AA\ and $\approx 143$\,\AA, respectively.

Turning our attention to comparable observations in the local Universe, \citet{sen19} and \citet{ber19} reported the detection of strong \civ\ emission with \ew(\civ) $\approx3-10$\,\AA\ from two low-mass, metal-poor galaxies at $z\sim0$ that also show strong \heii\ emission. Strong and narrow \heii\ emission has also been detected in these galaxies with \ew(\heii) $\approx 2.8$. Both studies concluded that stellar metallicities less than 10\% solar \citep[e.g.][]{sen21} are required to explain this emission, and that these galaxies are likely analogues of metal-poor star-forming systems in the reionisation era. 

Interestingly, rest-frame UV spectroscopy of LyC leakers at $z<0.7$ has also resulted in the detection of strong \civ\ emission in all galaxies with \fesc\,$>0.1$ \citep{sch22}, with some of the highest \ew(\civ) seen in low-$z$ star-forming galaxies. The \ew(\civ) for a majority of their LyC leaking sources are comparable to that of our stacked spectrum.

Overall, the properties of emission lines in the stacked spectrum of \civ\ emitters presented in this study appear to be consistent with those seen in strong LAEs at $z\sim3-5$, and are somewhat representative of the limited detections of rest-UV emission from bright galaxies at $z>6$. The line strengths in our sample also resemble those from extremely low-metallicity galaxies in the local Universe, as well as strong LyC leakers at $z\lesssim 1$.

\section{Underlying sources of ionising photons}
\label{sec:ionisation}
Having produced a stacked spectrum of strong \civ\ emitters and identified prominent rest-frame UV lines, in this section we compare the properties of the stacked spectrum with photoionisation models in a bid to constrain the dominant mechanism of ionising photon production within \civ\ emitting galaxies. 

\subsection{Spectral energy distribution (SED) fitting}
\label{sec:SED}
To understand the nature of ionising sources that could give rise to strong \civ\ emission (as well as other lines), we now find the best-fitting SED model for our stacked spectrum. Since the main aim of fitting SEDs to our stacked spectrum of \civ\ emitters is to broadly investigate what physical conditions within galaxies may give rise to strong \civ\ (and other line) emission, we choose to compare with fairly simple stellar population synthesis models, opting to use SEDs containing stellar continuum, nebular continuum and nebular line emission produced by the BPASS team \citep[e.g.][]{xia18}. These models are built on stellar SEDs generated using BPASS v2.2.1 for a single age starburst, with stellar ages in the range $\log(\rm{age/yr}) = 6 - 7.5$ \citep{xia18, sta18}. These SEDs are generated using the default BPASS initial mass function (IMF), which is a broken power law with a slope of $-1.30$ for $M_\star < 0.5 M_\odot$ and a slope of $-2.35$ for $M_\star > 0.5 M_\odot$, with an upper mass cutoff at $300 M_\odot$.

These simple stellar populations models are then processed using the photoionisation code Cloudy, assuming a nebular gas cloud with density $\log(n_H/\rm{cm}^{-3}) = 2.3$, a spherical geometry and the same metallicity of stars and nebular gas to calculate nebular line fluxes\footnote{\url{https://flexiblelearning.auckland.ac.nz/bpass/4.html}}. For this analysis, we consider Cloudy processed models containing stellar and nebular emission with stellar metallicities in the range $Z=0.005\,Z_\odot$ to $Z=2\,Z_\odot$ and the dimensionless ionisation parameter, $\log(U)$\footnote{SED models often tend to employ the dimensionless ionisation parameter, $U$, which gives the ratio of the density of ionising photons to the density of hydrogen. The ionising photon production efficiency or \xiion, on the the other hand is the number of ionising photons produced per unit UV luminosity. Under assumptions of hydrogen density for a given UV luminosity, $U$ and \xiion\ are closely related to one another and broadly trace the ionising photon production capabilities of a source.}, in the range $[-1.0, -4.0]$. We additionally include a dust attenuation prescription derived by \citet{red16b} for $z\sim3$ galaxies at rest-frame UV wavelengths, with the same attenuation applied to both nebular lines and the stellar continuum, considering $E(B-V)$ values in the range $[0,1]$ in steps of $0.01$.
\begin{figure*}
    \centering
    \includegraphics[width=\textwidth]{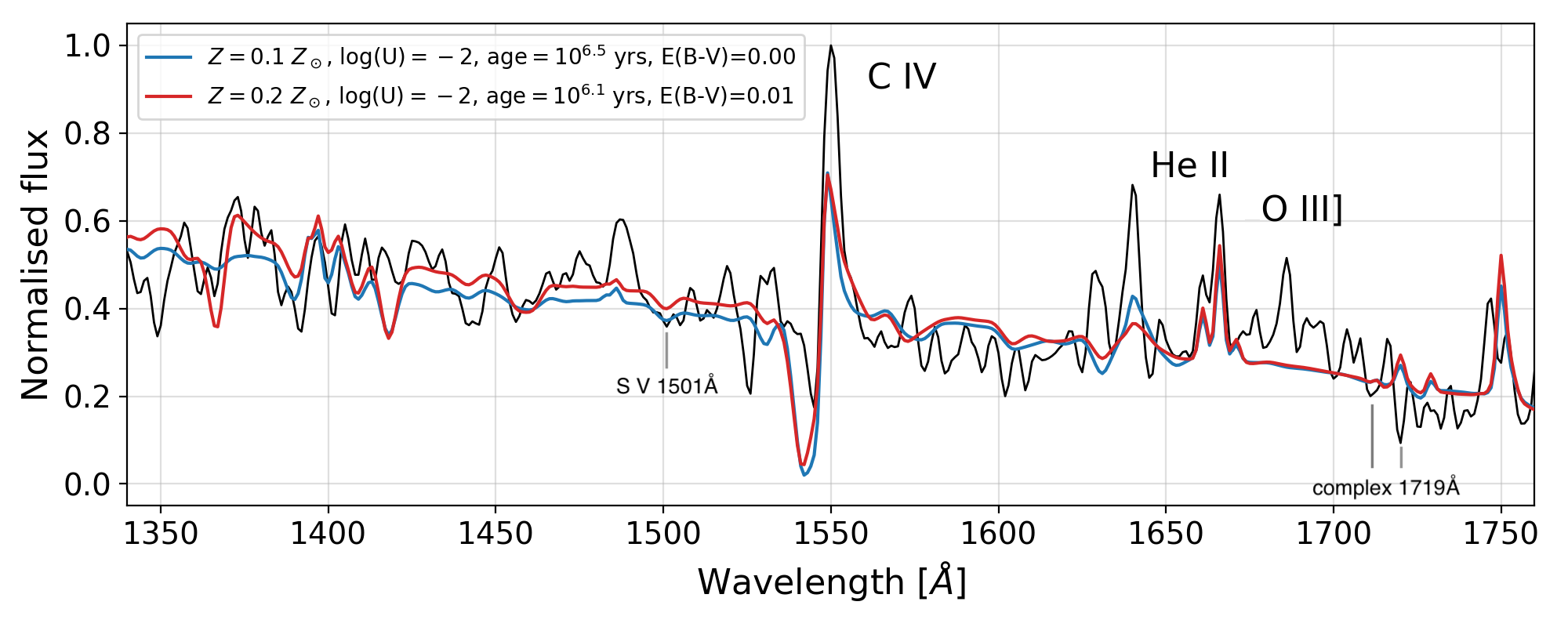}
    \caption{Zoom-in of the wavelength range covering \civ, \heii\ and \oiiiuv\ emission lines in the stacked spectrum of \civ\ emitters. The spectrum has been normalised using minimum-maximum scaling to aid comparisons with SED models (see text in \S\ref{sec:SED}), which have also been smoothed to the VANDELS spectral resolution. We show the two best-fitting spectral energy distribution (SED) models. Model 1 (blue) has a stellar metallicity of $Z=0.1\,Z_\odot$, $\log(\rm{age/yr}) = 6.5$, an ionisation parameter of $\log(U) = -2.0$ and no dust attenuation. Model 2 (red) has a higher stellar metallicity of $Z=0.2\,Z_\odot$, a lower stellar age of $\log(\rm{age/yr}= 6.1$, an ionisation parameter of $\log(U) = -2.0$ and a small dust attenuation of $E(B-V)=0.01$. We find that both models are able to reproduce the P-Cygni profile of \civ, but under-predict the \civ\ and \heii\ lines. The models match the \oiiiuv\ doublet strength as well as their ratio reasonably well. Overall, the best-fitting SED models suggest that young and low-metallicity starbursts with a high ionisation parameter are needed to explain the observed \civ\ (and \heii) in the stacked spectrum. We additionally mark the locations of the absorption features used to independently measure stellar metallicities in \S\ref{sec:metallicity}: S\,\textsc{v} at $1501$\,\AA\ and a blend of N\,\textsc{iv}, Si\,\textsc{iv}, Al\,\textsc{ii} and Fe\,\textsc{iv} at $1719$\,\AA. Finally, we note that the emission feature at $1486.5$\,\AA\ is from N\,\textsc{iv}] and the feature spotted near $\approx1680$\,\AA\ is likely a noise fluctuation.}
    \label{fig:best-fit-model}
\end{figure*}

Since the \lya\ line is prone to high levels of scatter and due to the increased noise levels at the edges of the stacked spectrum, we focus our model comparison on the wavelength range $1300-1820$\,\AA\ that contains \civ, \heii\ and \oiiiuv\ lines. It is important to match the normalisation of observed spectra with synthetic SEDs generated by models to enable accurate comparison. To this end, in this study we employ the \emph{MinMaxScaler} routine \footnote{\url{https://scikit-learn.org/stable/modules/generated/sklearn.preprocessing.MinMaxScaler.html}}, which scales and translates the features in a given array on to a finite range, in this case between zero and one. When this scaling is applied consistently across the observed spectrum and the model SEDs, an accurate comparison between data and models becomes possible.

To find the best-fitting SED to our stacked spectrum, we employ the root-mean-square deviation (RMSD) estimator. The RMSD estimator measures the differences between values predicted by a model and data, where the deviations represent the residuals. Low RMSD values indicate the closest agreement between models and data. We populate a grid of models with varying combinations of metallicities, ages, ionisation parameters and dust attenuation in the ranges described above, calculating the RMSD for each model compared to the stacked spectrum. We find that two SEDs in particular give comparably good fits to the stacked spectrum, having the lowest RMSD values, which are described below and shown in Figure \ref{fig:best-fit-model}. 

We note that when fitting SEDs we do not mask faint ISM absorption lines, which are not present in the models that we use. Owing to the presence of strong emission lines and relatively weak ISM absorption, we find that the principal components driving the model fit to the data are the multiple strong emission lines in the stacked spectrum, followed by the UV slope that covers a broad range of wavelengths.

The first model (Model 1) has a stellar metallicity of $Z=0.1\,Z_\odot$, a stellar age of $\log(\rm{age/yr}) = 6.5$, an ionisation parameter $\log(U) = -2.0$ with no dust attenuation, $E(B-V)=0.0$, shown in blue in Figure \ref{fig:best-fit-model}. Model 1 broadly reproduces the \civ\ emission line's P-Cygni profile but under-predicts the line flux and over-predicts the absorption bluewards of the peak. The \heii\ emission in the model is also not enough to match the observed line emission, however the \oiiiuv\ doublet line strengths and ratios, as well as \niii\ emission are well reproduced, with a good match to the observed UV slope of the stacked spectrum.

The second model (Model 2) has a higher stellar metallicity of $Z=0.2\,Z_\odot$ compared to Model 1, with a lower stellar age of $\log(\rm{age/yr}) = 6.1$, $\log(U) = -2.0$ and a small dust attenuation of $E(B-V)=0.01$, shown in red in Figure \ref{fig:best-fit-model}. The \civ\ line profile in Model 2 is nearly identical to Model 1, under-predicting the emission but over-predicting the absorption. The \heii\ emission in Model 2 is lower when compared to Model 1, but the \oiiiuv\ and \niii\ emission are well reproduced. Model 2 also matches the observed UV slope in the spectrum. Both Models 1 and 2 have highly comparable RMSD values, and their properties are summarised in Table \ref{tab:SED}.
\begin{table}
\centering
\caption{Properties of the best-fitting BPASS+Cloudy SED models}
\begin{tabular}{l c c c c c}
    \hline
     & $Z / Z_\odot$ & log(age/yr) & $\log(U)$ & $E(B-V)$ & RMSD \\
    \hline
    Model 1 & 0.1 & 6.5 & $-2.0$ & 0.00 & $0.098$ \\
    Model 2 & 0.2 & 6.1 & $-2.0$ & 0.01 & $0.099$ \\
    \hline
\end{tabular}
\label{tab:SED}
\end{table}

Both well-fitting SEDs imply that stellar populations with metallicities of $0.1-0.2\,Z_\odot$, low stellar ages and relatively high ionisation parameters are needed to reproduce the vast majority of the observed rest-UV emission lines and match the UV slope of the stacked spectrum of \civ\ emitting sources. Interestingly, the largest discrepancy between observations and model predictions are for emission lines requiring higher ionisation energies. For example, the \oiiiuv\ and \niii\ lines that require energies of 35.1 eV and 29.6 eV, respectively, are relatively well produced by both models, but the \civ\ and \heii\ lines requiring energies of 47.9 eV and 54.4 eV are under-predicted. We note here that further increasing the ionisation parameter of these models results in a catastrophic mismatch between the observed and predicted UV slopes, which is not significantly improved by increased dust attenuation, leading to sub-optimal fits. 

The inability of these models to accurately match the observed \civ\ and \heii\ emission suggests that the best-fitting SEDs are not producing ionising radiation fields that are `hard' enough, such that a large amount of photons with extremely high energies are produced, which we discuss further in \S\ref{sec:heii-civ}. Further, the observed \civ\ absorption blueward of the peak is much weaker in the stacked spectrum compared to what both models predict, indicating that the best-fitting models may be more metal rich than the observations might suggest. Therefore, to obtain constraints on stellar metallicities we use additional spectroscopic indicators in the following section.

\subsection{Stellar metallicity from absorption indices}
\label{sec:metallicity}
We now also measure the stellar metallicity of the stacked spectrum in a more direct manner using rest-frame UV absorption lines. We note the presence of absorption features due to ionised S\,\textsc{v} at $\approx1501$\,\AA\ and due to a blend of N\,\textsc{iv}, Si\,\textsc{iv}, Al\,\textsc{ii} and Fe\,\textsc{iv} at $\approx1719$\,\AA\ in the stack. These features arise as a result of absorption in the stellar photospheres of young, hot stars, and the strength of absorption can be used as a reliable tracer of stellar metallicity, independent of age of the stars or their initial mass function \citep[e.g.][and references therein]{cal21} assuming a constant star-formation history. 

We note that these calibrations may need to be revised for other star-formation histories, such as those predicting a larger fraction of older stars. However, given the smaller timescales involved when modelling galaxies at $z>3$ older stars may not play a big role \citep[see][for example]{cul19}. Another possible source of uncertainty is the spectral resolution of the observations, which, as \citet{cal21} showed can introduce errors of up to 0.2\,$Z_\odot$ in the inferred metallicities. Encouragingly, the spectral resolution used to calibrate the metallicity indicators in \citet{cal21} are at the VANDELS resolution, thereby minimising its effects in this study.

We measure \ew(1501)\,$=-1.0 \pm 0.5$\,\AA\ and \ew(1719)\,$=-1.3 \pm 1.0$\,\AA\ and using the metallicity calibrations from \citet{cal21} that use BPASS models without any nebular emission, we obtain a stellar metallicity of $Z = 0.0036 \pm 0.0025$ from the $1501$ index and a comparable $Z = 0.0041 \pm 0.0031$ from the $1719$ index. Using the solar metallicity value of $Z_\odot = 0.02$ as before, both these independent metallicity measurements from best-fits imply $Z \approx 0.2\, Z_\odot$ within the uncertainties. 

The best-fitting metallicity measurement is consistent with Model 2 presented in \S\ref{sec:SED}, although given the large uncertainties on the measurement a lower metallicity of $\approx0.1\,Z_\odot$ suggested by Model 1 is also possible. From the best-fits it appears that galaxies with strong \civ\ emission requiring high-energy ionising photons do not necessarily have abnormally low metallicities when compared with measurements for galaxies at $z\sim3.5$ from VANDELS, which have stellar metallicities in the range $Z\approx0.1-0.2\,Z_\odot$ \citep{cul20, cal21}. The picture is different in the local Universe, however, where known \civ\ and \heii\ emitters almost always have extremely low stellar (and gas phase) metallicities of the order $Z\lesssim0.1\,Z_\odot$ \citep[e.g.][]{ber19, sen21}. However, given the large uncertainties as well as the considerably stronger \civ\ absorption features in the best-fitting SEDs compared to observations, a very low metallicity solution for \civ\ emitters cannot be fully ruled out.

\begin{figure}
    \centering
    \includegraphics[width=\linewidth]{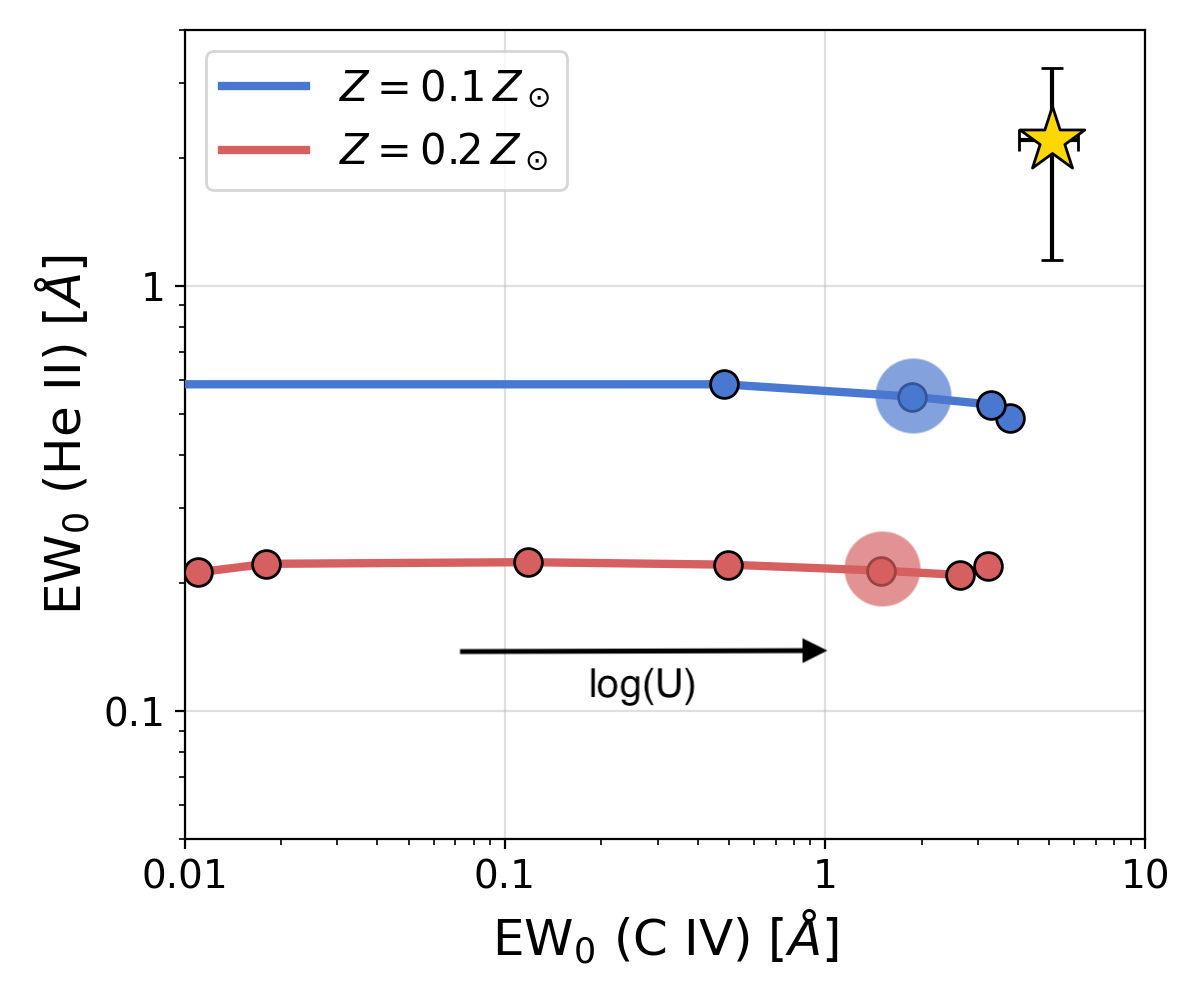}
    \caption{Comparison of the \ew(\civ)\ and \ew(\heii) measured in the stacked spectrum (gold star) with predictions from best-fitting models: Model 1 with $Z=0.1\,Z_\odot$, $\log(\rm{age/yr}) = 6.5$ (blue) and Model 2: $Z=0.2\,Z_\odot$, $\log(\rm{age/yr} = 6.1$ (red). The best-fitting ionisation parameter values of $\log(U) = -2$ for both models are highlighted. Particularly for Model 2, which has a consistent metallicity with that inferred from absorption features, the \heii\ line strength is under-predicted by an order of magnitude and the \civ\ line strength by a factor of $\sim3$. Additional sources of high-energy ionising photons may be needed to explain the observed emission line strengths.}
    \label{fig:ew-compar}
\end{figure}

\subsection{Under-prediction of \civ\ and \heii}
\label{sec:heii-civ}
We note again that neither of the two best-fitting SEDs presented in \S\ref{sec:SED} is able to fully reproduce the observed \civ\ and/or \heii\ emission in the stacked spectrum. This is also shown clearly in Figure \ref{fig:ew-compar}, where the ionisation parameter increases from right to left, with the highest value $\log(U)=-1.0$. The best-fitting $\log(U)$ value of $-2$ obtained from SED fitting has been highlighted for both models. 

Looking at Model 2 with $Z=0.2\,Z_\odot$, the \heii\ \ew\ is under-predicted by an order of magnitude at $\gtrsim 3\sigma$ significance, whereas the \civ\ line is under-predicted by a factor of $\sim3$, also with $\gtrsim 3\sigma$. The discrepancy between observed and predicted equivalent widths also exists when comparing with Model 1, but Model 1 produces stronger \heii\ in comparison to Model 2, bringing the tension down to $\sim 2\sigma$. The discrepancy with the \civ\ line, however, remains at $\sim 3\sigma$. 

An under-prediction of \heii\ equivalent widths even by the lowest metallicity models from BPASS was previously also reported by \citet{sax20}. To account for the `missing' \heii\ ionising photons within such galaxies, \citet{sax20} suggested the inclusion of additional sources of ionisation such as faint AGN, stripped binary stars \citep[e.g.][]{got19} or ultra-luminous X-ray sources \citep[e.g.][]{sch19, sax20b, sim21, ume22}. One additional scenario that may explain strong \civ\ and \heii\ emission is a recent starburst event (preferentially from metal-poor gas) within galaxies that harbour relatively enriched ($Z\gtrsim 0.2\, Z_\odot$) stellar populations.

Evidence for the presence of both young and evolved stellar populations in high-redshift galaxies that show strong emission lines has been recently discussed: broadband SED fitting and Atacama Large Millimetre Array (ALMA) observations of hyperfine transition metal lines of galaxies at $z\gtrsim9$ have suggested that these may already harbour evolved stellar populations \citep[e.g.][]{rob20, lap21}. Recently, \citet{tan22} also reported the presence of evolved stellar populations in extreme \oiii\,$\lambda5007$ emitting galaxies at redshifts $1.3-3.7$, where this evolved population can be up to $\sim40$ times more massive than the young starburst associated with the extreme line emission.

Therefore, galaxies undergoing periods of starburst activity may temporarily be able to produce copious amounts of high-energy photons (i.e., periods of high \xiion), giving rise to strong emission lines such as \civ, and \heii, as well as \oiii\,$\lambda5007$. Radiative transfer calculations in zoom-in simulations of galaxy formation from \citet{bar20} showed that periods of high \xiion\ are often coincident with periods of extreme emission line strengths driven by starburst events. Periods of high LyC \fesc\ tend to then follow, once the gas has been blown out and channels of low absorption have been established over a timescale of $\sim$Myrs due to rampant supernova activity. 

Having seen possible evidence of elevated ionising photon production in \civ\ emitting galaxies, in the next section we investigate whether \civ\ emitters may also trace conditions that enable a higher fraction of LyC photons to escape into the IGM to assess whether \civ\ emission in the reionisation epoch may effectively trace LyC leaking galaxies.

\section{C\,\textsc{iv} emitters as tracers of high ionising photon escape}
\label{sec:fesc}

The primary goal of this paper is to investigate whether \civ\ emitting galaxies at intermediate redshifts trace conditions that might enable a high LyC \fesc, which we test in this section using spectroscopic indicators in the stack.

\subsection{Inference from \lya\ strength and velocity offset}
\label{sec:lya}
Useful information regarding the presence of channels in the ISM, allowing for a high escape fraction of LyC photons, can be obtained from both the strength \citep[e.g.][]{dij14}, the profile and the velocity offset compared to systemic of the \lya\ emission line \citep[e.g.][]{ver15}. The redshift range of our targets ensures that the \lya\ line is visible for all galaxies, and in this section we use the observed \lya\ emission in the stacked spectrum to explore the possibility of high LyC \fesc\ from \civ\ emitting galaxies.

It has been shown that galaxies with high LyC \fesc\ are expected to have the peak of their \lya\ emission line close to the systemic redshift, with non-zero \lya\ flux bluewards of the systemic redshift \citep{ver15, dij16}. In this scenario, both \lya\ and LyC photons would be able to escape the H\,\textsc{ii} regions within which they are produced with relative ease through a porous ISM. This has also been verified observationally through \lya\ profiles of low redshift LyC leakers, with an anti-correlation observed between LyC \fesc\ and the separation of the \lya\ blue and red peaks as well as velocity offset from systemic \citep[e.g.][]{izo18a, izo21}.  

\begin{figure}
    \centering
    \includegraphics[width=\linewidth]{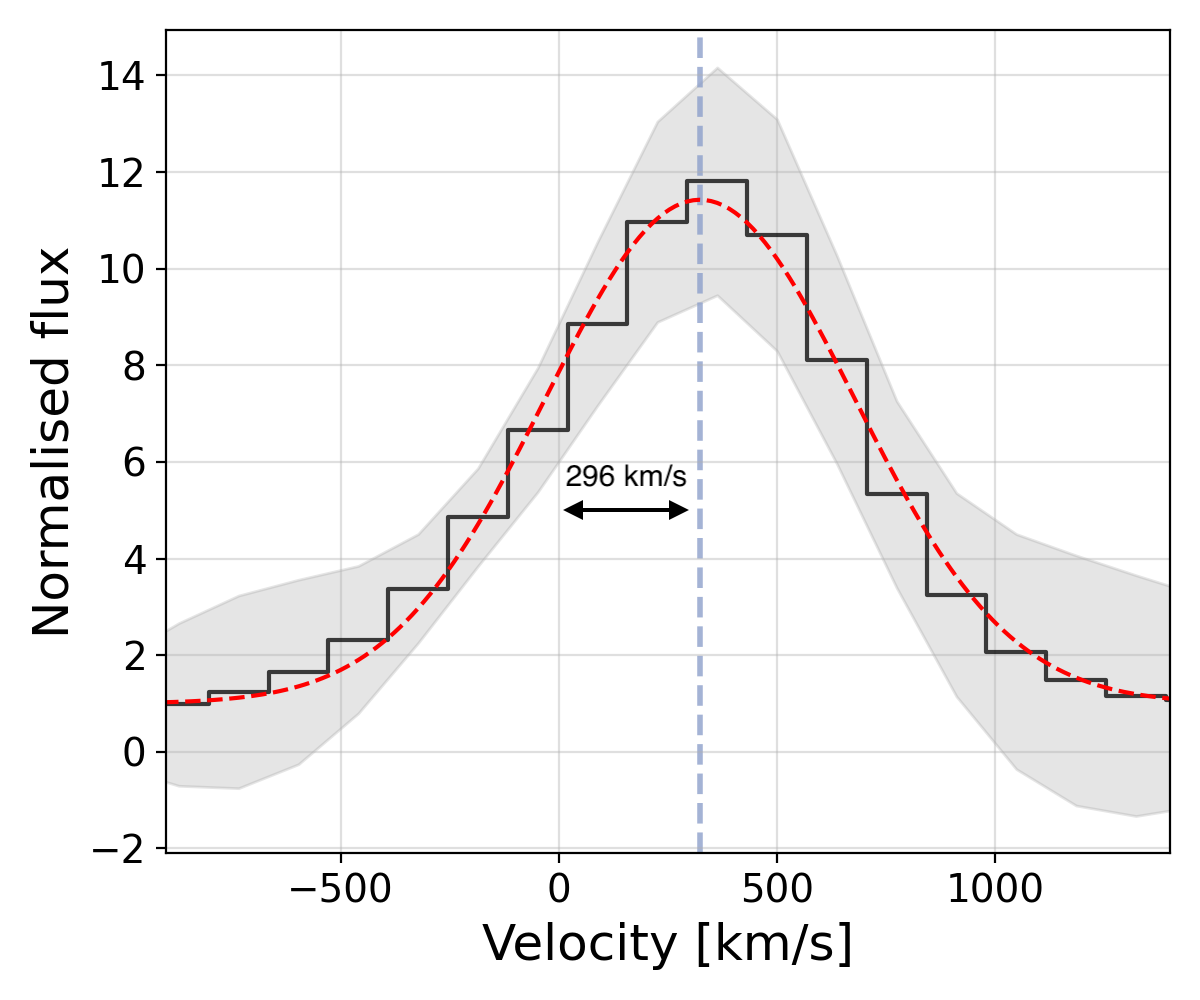}
    \caption{Zoom-in of the \lya\ emission line in the stacked spectrum in units of velocity, with $v=0$ tracing the systemic velocity constrained using the \civ\,$\lambda1550$ feature in the stack. The velocity offset of the \lya\ peak is $\approx300$\,km\,s$^{-1}$ compared to the systemic velocity and there is significant flux bluewards of the \lya\ peak. The line profile is highly symmetric. All of these \lya\ features are consistent with expectations from galaxies that may or do indeed show a high escape fraction of LyC photons \citep[e.g.][]{ver15, ver17}.}
    \label{fig:lya-vel}
\end{figure}

We find that the velocity offset of the \lya\ line in the stacked spectrum compared to the systemic redshift is relatively small at $v\approx296 \pm 20$\,km\,s$^{-1}$, as shown in Figure \ref{fig:lya-vel}. We also note that the \lya\ emission is strong and appears to be symmetric, reproduced well by a single Gaussian function with non-zero flux bluewards of the \lya\ peak (see also \S\ref{sec:stack-lines}). However, the emergent symmetrical profile in the stacked spectrum may be a consequence of stacking objects with a variety of \lya\ profiles and velocity offsets, with low spectral resolution additionally resolving out any intrinsic multi-peak structure of \lya\ emission seen in individual objects. The strength and the low velocity offset compared to systemic, however, are consistent with predictions from \citet{ver15} of high \lya\ and LyC photon escape.

To put the observed \lya\ strength and velocity offset observed in the stacked spectrum of \civ\ emitters in the global context of star-forming galaxies at $z\sim3.6$, we compare our measurements to stacked spectra of 19 randomly selected star-forming galaxies from the VANDELS parent sample with the same range of UV luminosities, redshifts and redshift quality flags as the \civ\ emitters. We repeat the stacking process 500 times following the methodology outlined in \S\ref{sec:stack}, obtaining a distribution of stacked spectra of non-\civ\ emitters randomly drawn from VANDELS. We then measure the strength of \lya\ emission in each stack as well as the offset of the peak of \lya\ emission from the expected `systemic' redshift, using the spectroscopic redshifts compiled by the VANDELS team as reference.
\begin{figure}
    \centering
    \includegraphics[width=\linewidth]{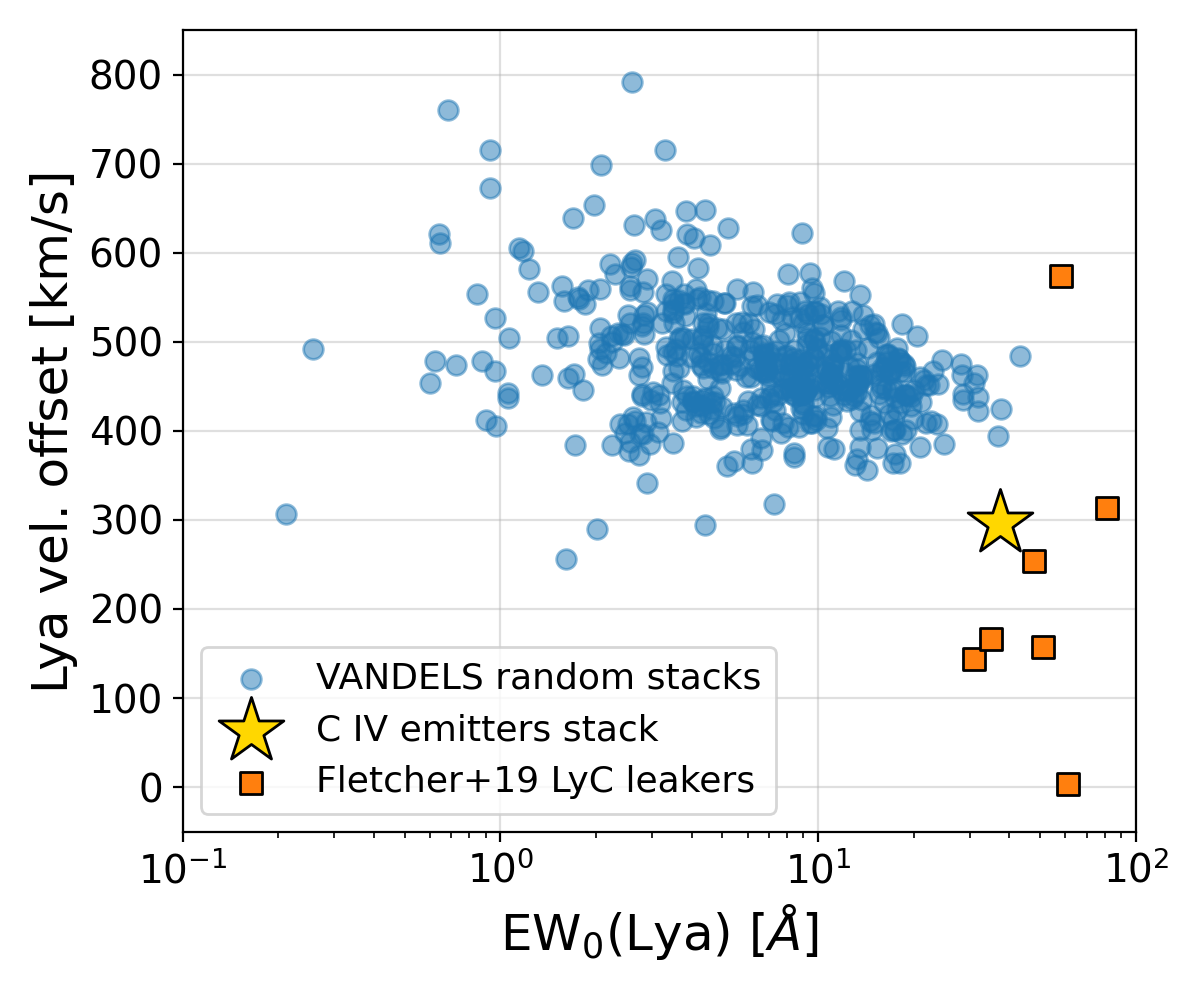}
    \caption{Distribution of \lya\ EW$_0$ and velocity offset from systemic redshift measured from the stacked spectrum of \civ\ emitters, compared with measurements from 500 iterations of stacking 19 randomly drawn galaxies from the VANDELS survey occupying the same redshift range as the \civ\ emitters. The \lya\ strength and velocity offset of \civ\ emitters is an outlier compared to the distribution inferred from randomly selected VANDELS galaxies, implying that the \lya\ emission line from \civ\ emitting galaxies is indicative of high LyC \fesc\ \citep[e.g.][]{ver15}. We additionally show \lya\ measurements from LyC leakers at $z\sim3$ presented in \citet{fle19}, which also exhibit strong \lya\ as well as low velocity offset from the systemic redshift.}
    \label{fig:lya-ew-vel}
\end{figure}

We show the results of this exercise in Fig \ref{fig:lya-ew-vel}, where it is clear that the stack of \civ\ emitters not only shows a \ew(\lya) value that is $5\sigma$ higher than the median \lya\ EW measured from randomly stacking VANDELS spectra, its velocity offset compared to systemic is also $3\sigma$ lower than the median offset. We also find that stronger \lya\ lines tend to peak closer to the systemic redshift, which has also been observed in the literature \citep[see][for example]{erb14}. 

Also shown in Figure \ref{fig:lya-ew-vel} are \lya\ measurements from LyC leakers identified by \citet{fle19} from a sample of narrow-band selected LAEs, which have comparable \lya\ line strengths and offsets from the systemic velocity to those seen in our stacked spectrum\footnote{Here we use all LyC leakers from \citet{fle19} that have \ew(\lya) measurements from spectroscopy, with accurate systemic redshifts from other rest-UV lines. These galaxies belong to both \emph{Gold} (4 galaxies) and \emph{Silver} (3 galaxies) sub-samples.}. This suggests that galaxies with strong \civ\ emission preferentially show stronger \lya\ emission peaking close to the systemic redshift, as is often seen in the spectra of LyC leaking galaxies across redshifts.

Other known LyC leaking galaxies at high redshifts in the literature show similar \lya\ strengths and profiles, with \lya\ from the \emph{Sunburst Arc} at $z=2.4$ \citep{riv19}, \emph{Ion2} at $z=3.2$ \citep{van16a} and \emph{Ion3} at $z=4.0$ \citep{van18} peaking close to systemic $v=0$ with non-zero flux bluewards of the peak. High resolution spectra for such galaxies have revealed multiple peak \lya\ morphologies, indicative of ionising photon escape channels in the neutral H\,\textsc{i} gas \citep[e.g.][]{van18}. 

The \lya\ line seen in our stacked spectrum is also comparable to that of the stacked spectrum of the subsets of LAEs at $z\sim2$ that are likely leaking significant LyC radiation presented by \citet{nai22}. Those authors also find \civ\ emission in the stacked spectrum of candidate LyC leakers, with no \civ\ emission detected in the stack of LAEs that are unlikely to be leaking LyC photons. 

We note here that 3 out of 19 \civ\ emitters in our final sample do not show any \lya\ emission in their spectra, which is reminiscent of no \lya\ emission being observed in the spectrum of a strong LyC leaker \emph{Ion1} at $z=3.79$ \citep[][and references therein]{ji20}. Finding no \lya\ emission in a LyC leaking galaxy is puzzling, but our current understanding of the connection between \lya\ strength and LyC leakage, especially at intermediate redshifts, is also based on very limited samples. A likely explanation of the absence of \lya\ in the presence of other strong UV lines and LyC leakage is that strong \lya\ and LyC leakage may not be coincident \citep[see][for example]{kee17}, which is especially true for galaxies with clumpy morphologies \citep[e.g.][]{riv19}, or the possibility of resonant scattering removing \lya\ from our direct line-of-sight \citep{ji20}.

Due to the relatively low spectral resolution of our stacked spectrum we are unable to resolve the \lya\ line to explore in more detail indication of significant LyC \fesc. However, future higher resolution observations of the \lya\ line profile of \civ\ emitting galaxies would be valuable to further investigate the presence of \lya/LyC escape channels.

\subsection{Presence of strong He\,\textsc{ii}}
\label{sec:heii}
We also find strong \heii\ emission in addition to \civ\ emission in the stacked spectrum, with \heii/\civ\,$\approx0.5$, indicative of the presence of sources capable of producing hard ionising radiation fields \citep[see also][]{ber18, ber19}. \citet{sch22} also showed the presence of \heii\ emission in strong LyC leaking galaxies at $z<0.7$, with high equivalent widths of $3-8$\,\AA, which can be explained with relatively high ionising photon production efficiencies, log(\xiion)\,$\approx25.6-25.8$\,erg$^{-1}$\,Hz. These line strengths are higher than \ew(\heii) $= 2.6$ that we measure from our stacked spectrum, but the relatively high \heii/\civ\ ratio we find is comparable to that seen in the spectra of LyC leakers presented by \citet{sch22}, indicative of similarly high \xiion.

\begin{figure}
    \centering
    \includegraphics[width=\linewidth]{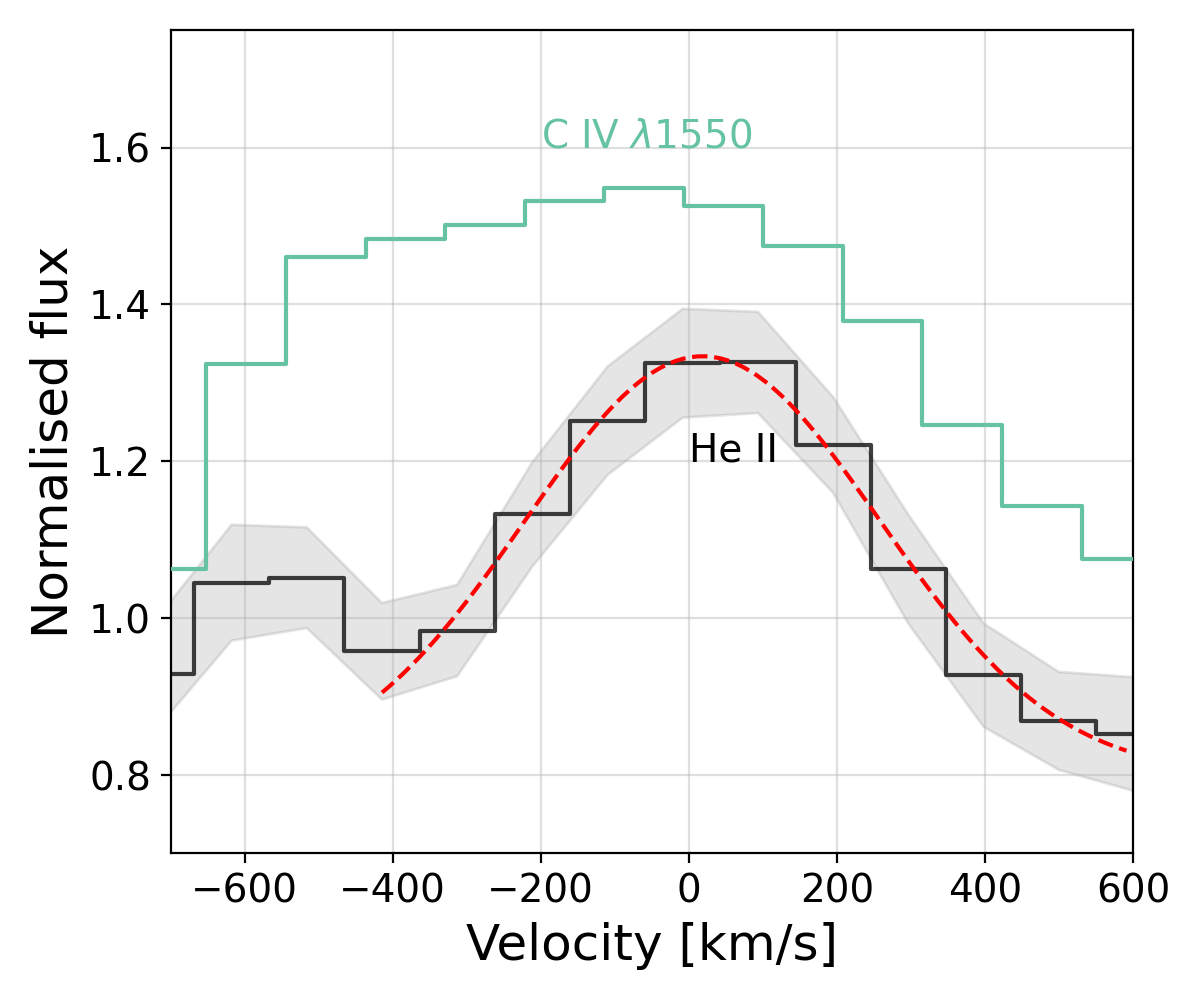}
    \caption{Velocity profile of \heii\ emission (black) in the stacked spectrum shown alongside the \civ\ emission (green), with $v=0$ tracing the systemic velocity. The \heii\ line is well-fit with a single Gaussian (red dashed line) with the peak of the Gaussian within $\approx50$\,\kms\ of the systemic velocity. The presence of both \civ\ and \heii\ emission lines requires the production of extremely high-energy photons ($E>47.9$ and $E>54.4$ eV for \civ\ and \heii, respectively). We find a relatively high \heii/\civ\ ratio of $\approx0.5$, which is indicative of extremely high ionising photon production efficiencies of log(\xiion)\,$\gtrsim 25.6$\,erg$^{-1}$\,Hz \citep[e.g.][]{ber19, sch22}.}
    \label{fig:heii-civ}
\end{figure}

In Figure \ref{fig:heii-civ} we show the velocity profile of the observed \heii\ emission with respect to the \civ\ line, finding that the peak of \heii\ emission is within $\approx50$\,\kms\ to that of \civ, tracing the alignment of channels through which photons with much higher energies may also escape in addition to the relatively lower energy channels traced by \lya. With \civ\ being a resonant line, the peaks of \civ\ and \heii\ being coincident suggests that the resonant scattering does not dramatically alter the energies of escaping photons, pointing towards the presence of well-defined columns in the multi-phase ISM facilitating high-energy ionising (and possibly LyC) photon escape. 

Further, \citet{nai22} also found \heii\ emission in the stacked spectra of candidate LyC leaking LAEs at $z\sim2$ with \ew(\heii) $\approx2$\,\AA, which is highly comparable to our measurement. However, \citet{mar22} reported the detection of comparable \heii\,$\lambda4686$ emission from LyC leakers and non-leakers $z\sim0.2-0.4$, noting that metallicity and not LyC escape is the dominant factor in setting \heii\ line strengths. Their study implies that LyC leaking galaxies may not show systematically `harder' ionising spectra compared to non-leakers at similar metallicities, casting doubt on the presence of \heii\ emission as a standalone indicator of strong LyC leakage. Nonetheless, strong \heii\ emission in our stacked spectrum of \civ\ emitters, when combined with other indicators, may still favour high \fesc.

\subsection{Insights from low-ionisation interstellar absorption}
\label{sec:lis}
In this section we attempt to infer LyC \fesc\ from both the depths and the profiles of absorption features in the stacked spectrum of \civ\ emitters arising from singly ionised species in the ISM. A low covering fraction ($f_c<1$) of metal-enriched gas, and consequently of that of H\,\textsc{i} gas, is expected to be a necessary but not sufficient condition for significant LyC photon escape, and several studies have investigated this relationship using rest-UV absorption lines \citep{jon13, hen15, red16, lee16, red22, ste18, chi18, sal22}.

Since the wavelength range of our stacked spectrum does not cover the H\,\textsc{i} Lyman-series transitions of Ly\textbeta\ and beyond, we focus instead on other well-studied low-ionisation interstellar (LIS) absorption features at rest-UV wavelengths, C\,\textsc{ii}\,$\lambda1334$ and Si\,\textsc{ii}\,$\lambda1260$ and $\lambda1526$. We do not use the Si\,\textsc{ii}\,$\lambda1304$ feature as it appears to be contaminated by O\,\textsc{i}\,$\lambda1302$. 

Following \citet{jon13}, we calculate the covering fraction of the above mentioned transitions assuming a `picket fence' like distribution of optically thick gas clouds and optically thin `holes' in the ISM \citep[e.g.][]{ste18}. Within the gas clouds, the column density of gas is assumed to be high enough such that it appears optically thick ($\tau >>1$) at the absorbing wavelength, saturating the absorption lines. In this picture, the covering fraction may be calculated as $f_c = 1 - I(\lambda)/I_0$, where $I_0$ is the local continuum level and $I(\lambda)$ is the residual intensity in the spectrum \citep[see also][]{sal22}. We find an average covering fraction of LIS lines to be $\approx0.2$. Following the best-fit relation obtained by \citet{sal22} for the covering fraction of LIS lines and H\textsc{i}, we estimate H\textsc{i} covering fractions of $\approx0.7$ resulting in limits of LyC \fesc\,$\lesssim 0.3$.

We further measure \ew(LIS) following the methodology of \citet{sal22} by first locating the minimum depth of the absorption line in question and then integrating over a velocity range of $\pm1250$\,\kms, dividing the measured flux by the local stellar continuum. Since we use the convention whereby emission lines have positive equivalent widths, the equivalent width measured for absorption features in this work are negative. We find \ew(Si\,\textsc{ii}\,$\lambda1260$)\,$=-0.61$\,\AA, \ew(Si\,\textsc{ii}\,$\lambda1526$)\,$=-0.80$\,\AA\ and \ew(C\,\textsc{ii}\,$\lambda1334$)\,$=-0.69$\,\AA. 

Using the the average \ew(LIS) $\approx -0.70\pm0.8$\,\AA\ and the relation between \fesc\ and \ew(LIS) obtained by \citet{sal22}, the \fesc\ is estimated to be in the range $\sim0.05-0.30$. Finally, from only \ew(C\,\textsc{ii}\,$\lambda1334$) using the \citet{mau21} relation we infer \fesc\,$>0.1$. All of these estimates point towards significant \fesc\ from galaxies that show strong \civ\ emission in their spectra.

\begin{figure}
    \centering
    \includegraphics[width=\linewidth]{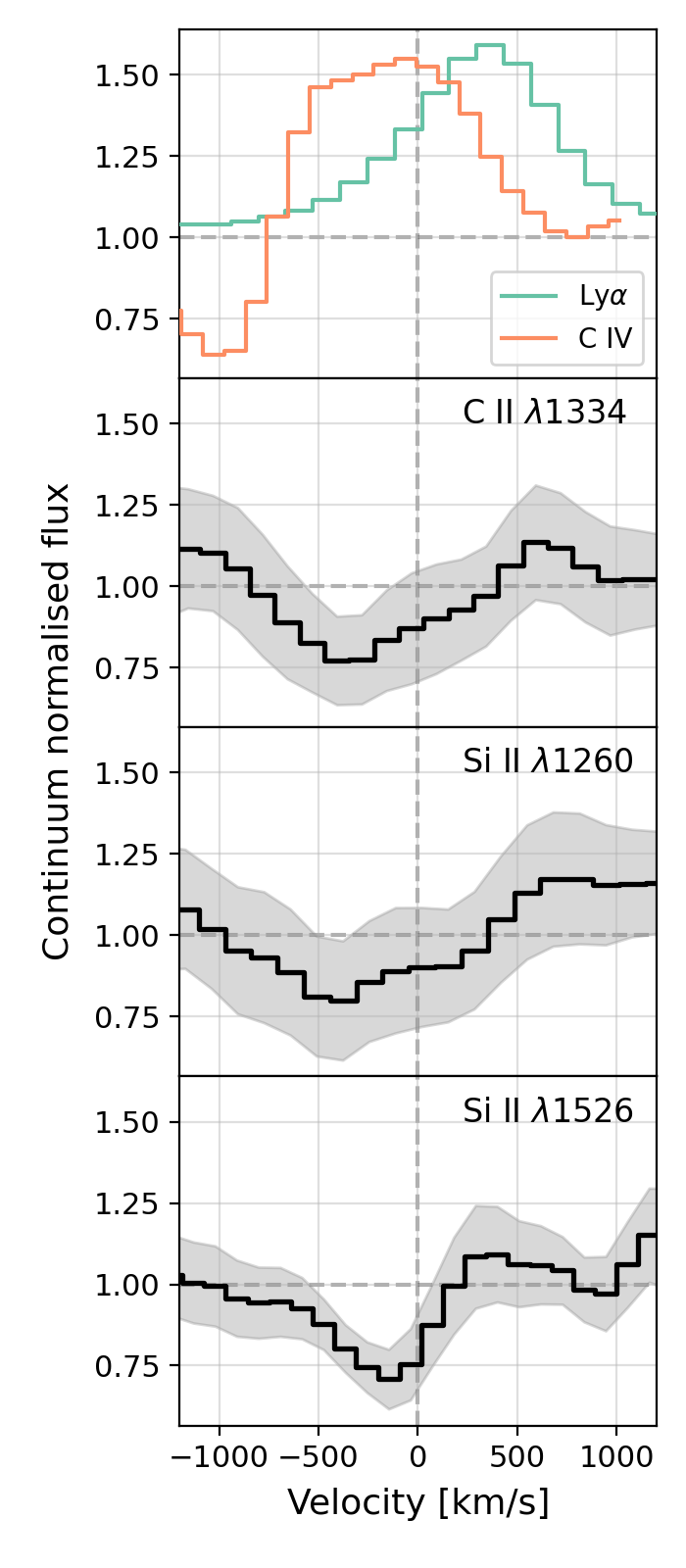}
    \vspace{-0.6cm}
    \caption{Low-ionisation interstellar (LIS) absorption of C\,\textsc{ii} $\lambda1334$, Si\,\textsc{ii} $\lambda1334$ and $\lambda1526$ shown along with strong \lya\ and \civ\ emission in the stacked spectrum. Here the continuum is set to 1 at $1500$\,\AA. The shaded regions around the LIS lines are $1\sigma$ uncertainties calculated using bootstrapping. The systemic velocity $v=0$ is set using \civ\,$\lambda1550$. All features have been continuum normalised, with the \lya\ line re-scaled for visualisation. The low covering fractions of LIS absorption features are indicative of relatively high \fesc\ \citep{mau21, sal22}, which also appear to be blue-shifted compared to the systemic velocity indicative of outflowing gas with velocities in the range $-500 \lesssim v \lesssim -200$\,km\,s$^{-1}$, consistent with measurements from known LyC leakers \citep[e.g.][]{chi17, ste18}.}
    \label{fig:absorption}
\end{figure}

These LIS absorption features along with strong \lya\ and \civ\ lines are shown in Figure \ref{fig:absorption}. Once again, $v=0$ is set to be at the peak of the \civ\,$\lambda1550$ emission. The shaded regions show $1\sigma$ uncertainties calculated using bootstrapping in \S\ref{sec:stack}. Here the flux density of the stacked spectrum is normalised to have a value of 1 at $1500$\,\AA, and the \lya\ emission has been re-scaled to aid visualisation.

We further note that the peaks of the absorption features are blue-shifted with respect to the systemic velocity, indicative of the presence of outflowing gas \citep[e.g.][]{jon13}. We infer outflow velocities in the range $-500 \lesssim v \lesssim -200$\,km\,s$^{-1}$, which are within one resolution element of the VANDELS spectra. The maximum covering fractions for the LIS gas are measured at $v\approx150-400$\,\kms.

It is important to note that the scatter on the relation between \fesc\ and \ew(LIS) can be large, and the relation is further sensitive to the spectral resolution as demonstrated by \citet{sal22}. Those authors showed that these effects may increase the error on the inferred covering fractions by $5-20\%$, with a compounded effect on the uncertainty on \fesc. We further note that stacking the spectra of individual galaxies with varying outflow velocities may also muddle the absorption troughs by artificially broadening the absorption features, which has an important implication that \fesc\ measured from the covering fraction of LIS lines from stacked spectra will be an upper limit. 

However, the widths of the LIS absorption features in the stacked spectrum appear to be fairly consistent with the spectral resolution of VANDELS, indicating that the individual outflow velocities do not vary dramatically across our sample. The outflow velocities for these lines that we infer are consistent with those measured for confirmed LyC leakers by \citet{chi17}. We further note the remarkable resemblance of both the absorption profiles as well as the outflow velocities with the high \fesc\ sub-samples from \citet{ste18}, who also reported an increase in the depth of these absorption features with decreasing LyC \fesc.

\subsection{C\,\textsc{iv}/C\,\textsc{iii}] ratios for individual galaxies}
\label{sec:c43}
In \S\ref{sec:lines} we noted the presence of \ciii\ emission in 5 galaxies in our sample, and fortunately all 5 of these sources are likely to be star-forming galaxies (and not AGN). \citet{sch22} recently reported the detection of both \civ\ and \ciii\ from a sample of confirmed LyC leaking galaxies at $z<0.7$, and found strong LyC leakers (\fesc\,$>0.1$) to have \civ/\ciii\ ratios in excess of 0.75. Therefore, in this section we explore whether strong LyC leakage can be inferred, at least qualitatively, from galaxies in our sample with both \civ\ and \ciii\ line detections.

Before comparing with measurements from low-$z$ leakers, we note that in the analysis of \citet{sch22} the \civ\ emission was likely purely nebular in origin, owing to a lack of P-Cygni absorption feature that is indicative of a stellar origin because of stellar photospheric absorption. We do note the presence of some absorption blueward of the \civ\ line both in individual sources and the stacked spectrum, and we now attempt to capture the fraction of \civ\ flux that could be attributed to stellar emission. To closely replicate the estimation of the stellar and nebular components in the \civ\ line in the \citet{sch22} analysis, we fit the individual galaxy spectra using SEDs that only contain stellar emission using the methodology of \citet{sal22}.

Briefly, these stellar-only SEDs use STARBURST99 single star models \citep{lei11} that include stellar rotation across a range of ages and metallicities. The individual spectra are converted to rest-frame and the model SEDs are convolved with a Gaussian kernel to match the VANDELS spectral resolution. Additionally employing a uniform foreground dust attenuation model, the observed spectra are fitted with a linear combination of STARBURST99 models (that only include light from the stellar continuum). We refer the readers to \citet{sal22} for more details about the fitting procedure. From this exercise, we find that on average $\approx25\%$ of the \civ\ flux may be attributed to stellar origin across our sample.

\begin{figure}
    \centering
    \includegraphics[width=\linewidth]{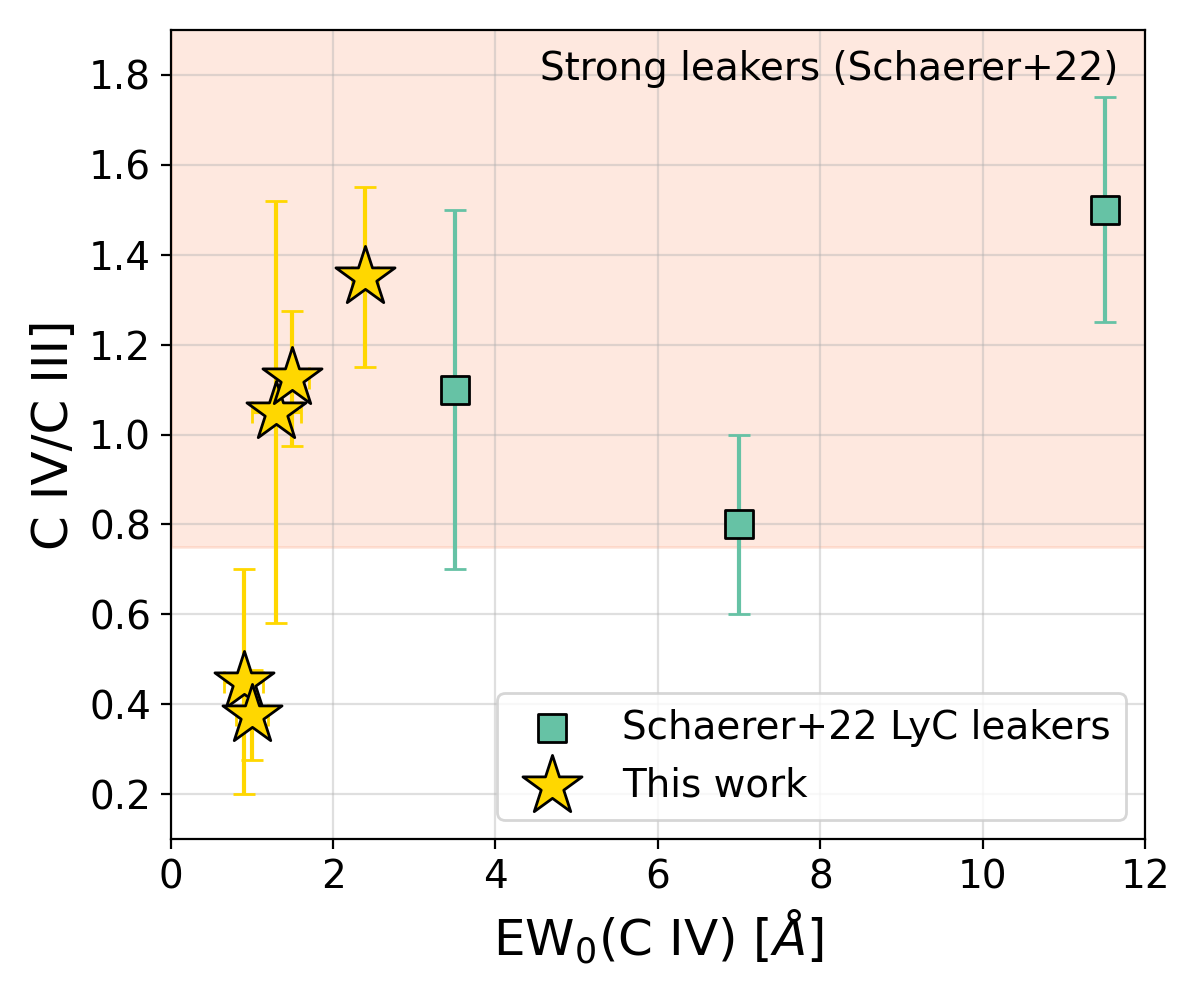}
    \caption{Ratio of nebular \civ\ and \ciii\ versus \ew(\civ) for individual galaxies in this study where both lines are reliably detected. Also shown for comparison are measurements for confirmed LyC leaking galaxies at $z<0.7$ from \citet{sch22}. The shaded area represents the parameter space where \citet{sch22} found LyC leaking galaxies in their sample to lie. We find that 3 out of 5 galaxies in our sample that have both \civ\ and \ciii\ detections show comparable \civ/\ciii\ ratios to that of LyC leakers from \citet{sch22}. We note that the \ew(\civ) for our galaxies are lower than that of low redshift leakers.} 
    \label{fig:civ-ciii-ratio}
\end{figure}

When plotting the \civ/\ciii\ ratio in Figure \ref{fig:civ-ciii-ratio}, we have removed the stellar contribution to \civ\ and only show the ratio of the nebular component of these lines as a function of \ew(\civ). We also show measurements from strong LyC leakers from \citet{sch22}. Interestingly, based on this simple diagnostic we infer that 3 out of 5 galaxies in our sample that show both these lines lie in the regime of strong LyC leakage. We do note that the strength of \civ\ emission from galaxies in our sample is systematically lower than what \citet{sch22} find for their LyC leakers.

Unfortunately, the \ciii\ line does not fall within the observed wavelength range of all galaxies in our final sample, and lying in a relatively redder part of the wavelength range of the spectrograph, the \ciii\ line is often contaminated by skyline residuals. Therefore, we choose not to consider \ciii\ emission measures drawn from our stacked spectrum. However, from a very simple \civ/\ciii\ ratio based diagnostic it is clear that strong LyC leakage may be expected from a fraction of \civ\ emitting galaxies in our sample, especially when considered in combination with other spectroscopic indicators in the stacked spectrum. 

\subsection{Concluding remarks on LyC escape}
We conclude this section by reiterating the numerous lines of evidence supporting the contention that galaxies selected by their strong \civ\ emission have ISM
conditions favourable for the escape of ionising LyC photons. These include strong \lya\ emission line peaking close to the systemic redshift with non-zero flux bluewards of the peak, the presence of high-ionisation lines such as \heii\ and low covering fractions of LIS absorption lines and their blue-shifted absorption troughs. We also find elevated \civ/\ciii\ ratios observed in individual \civ\ emitters where both lines are robustly detected, which is also expected under conditions of significant LyC leakage.

The remarkable presence of all of these features together strongly indicates, albeit indirectly, that \civ\ emitting star-forming galaxies may be good candidates of strongly LyC leaking galaxies. With NIRSpec on board \emph{JWST} it will be possible to detect \civ, \heii\ and \ciii\ emission from galaxies at $z>6$, while probing absorption features from metals as well as neutral hydrogen gas in galaxies at even higher redshifts, possibly providing reliable observables to infer LyC \fesc\ in the epoch of reionisation. We have shown that galaxies exhibiting strong \civ\ have both high \xiion\ as well as potentially high \fesc, the product of which is a key ingredient to understand the role of star-forming galaxies towards the cosmic reionisation budget at $z>6$.

\section{Summary}
\label{sec:summary}
In this work we have identified 19 \civ\ emitting star-forming galaxies from the VANDELS survey spanning a redshift range $z=3.1-4.6$ and presented their stacked spectrum (\S\ref{sec:data} and \S\ref{sec:stack}). Some individual \civ\ emitters show other rest-frame UV lines such as \heii\ and \oiiiuv\ along with \lya, and all of these lines are securely detected in the stacked spectrum allowing for a detailed analysis of the average properties of the underlying stellar populations as well as the interstellar medium in \civ\ emitting galaxies.

We show that the inferred rest-frame UV line fluxes and ratios from the stacked spectrum of \civ\ emitting galaxies at $z\sim3.6$ suggest that they are comparable in ionisation properties to local \civ\ (and \heii) emitting metal-poor galaxies, that have long been touted as analogues of reionisation era galaxies. We also find that the line strengths in the stack are similar to a handful of known \civ\ emitting sources at $z>6$ (\S\ref{sec:literature}).

For the stacked spectrum of \civ\ emitters, we find that the best-fit spectral energy distribution (SED) models incorporating both stellar continuum and nebular line emission have low stellar metallicities of $Z=0.1-0.2Z_\odot$, a young stellar ages of $\log(\rm{age/yr}) = 6.1-6.5$, a high ionisation parameter $\log(U)=-2.0$ and little to no dust ($E(B-V) = 0.00-0.01$). This suggests that the presence of young, metal-poor stellar populations is necessary to explain the strong \civ\ (and other rest-UV) line emission seen across our sample (\S\ref{sec:SED}).

We also measure the average stellar metallicity of \civ\ emitters from the stacked spectrum using absorption indices at 1501\,\AA\ and 1719\,\AA that are sensitive to the metal content of stars. Both indices give a stellar metallicity of $Z\approx0.2Z_\odot$ within errors, which is consistent with the the metallicity of SED model with $\log(\rm{age/yr}) = 6.1$ and a small dust attenuation of $E(B-V)=0.01$ (\S\ref{sec:metallicity}).

However, we find that the SED with $Z\approx0.2Z_\odot$ under-predicts the \civ\ equivalent width by a factor of 3, and the \heii\ equivalent width by a factor of 10. Since extremely young/metal-poor stars are needed to increase the predicted \civ\ and \heii\ line fluxes from models, the observations of these strong lines and relatively higher metallicities suggest that a relatively young starburst event within a galaxy containing older populations may be able to explain the observed spectral features of our stacked spectrum (\S\ref{sec:heii-civ}).

We then investigate whether galaxies showing strong \civ\ emission may exhibit significant hydrogen ionising LyC photon leakage into the intergalactic medium using a variety of indicators (\S\ref{sec:fesc}). First, we find that the strength and shape of the \lya\ emission in the stack is indicative of significant LyC leakage, as the \lya\ line peaks close to the systemic velocity and contains non-zero flux blueward of the peak (\S\ref{sec:lya}). The presence of strong \heii\ emission is indicative of substantial production of high-energy photons from young stars, and is consistent with what has been observed in the spectra of known LyC leaking galaxies (\S\ref{sec:heii}). The low equivalent widths and outflow velocities of low-ionisation interstellar absorption features are indicative of low column density channels through which LyC photons may escape, with their low covering fractions suggesting LyC \fesc\ $\approx 0.05-0.30$ (\S\ref{sec:lis}). Finally, the \civ/\ciii\ ratio of a fraction of \civ\ emitting galaxies is comparable to measurements from other known LyC leakers in the low redshift Universe (\S\ref{sec:c43}).

We therefore conclude that \civ\ emitting galaxies harbour young stellar populations, tracing recent starburst events that leads to the production of copious amounts of ionising photons. This starburst phase is needed to displace neutral as well as low-ionisation gas in the ISM of galaxies, potentially creating holes in the ISM through which LyC photons may be able to escape into the IGM. We find indirect evidence of significant LyC leakage from \civ\ emitting galaxies, suggesting that such galaxies could be important contributors towards cosmic reionisation at $z>6$. Conditions leading to strong \civ\ emission may be ubiquitous across galaxies at very high redshifts when the stellar populations were young and star-formation from metal-deficient gas was widespread. 

At $z\gtrsim6$ when the increased neutrality of the IGM attenuates \lya\ photons along the line-of-sight, the \civ\ line may offer a reliable alternative to identify galaxies with significant \fesc, which will be possible for statistical samples using \emph{JWST}/NIRSpec. However, careful radiative transfer modelling of \civ\ and LyC is needed to theoretically back up any dependence of \civ\ and \fesc. Therefore, the presence of strong \civ\ emission combined with other rest-UV indicators in the spectra of galaxies at $z>6$ can help establish whether they are likely contributors towards cosmic reionisation at early epochs.

\section*{Acknowledgements}
We thank the referee for constructive feedback that helped improve the quality of this work. AS thanks Silvia Genovese for assistance with plots and figures. AS and RSE acknowledge financial support from European Research Council Advanced Grant FP7/669253. ASL and DS acknowledge support from Swiss National Science Foundation.

This work makes use of VANDELS data. The VANDELS collaboration acknowledges the invaluable role played by ESO staff for successfully carrying out the survey. This work has made extensive use of Jupyter and IPython \citep{ipython} notebooks, Astropy \citep{astropy2}, hoki \citep{hoki}, matplotlib \citep{plt} and seaborn \citep{seaborn}. This work would not have been possible without the countless hours put in by members of the open-source developing community all around the world.

\section*{Data Availability}
The VANDELS Data Release 4 (DR4) is now publicly available and can be accessed using the VANDELS database at \url{http://vandels.inaf.it/dr4.html}, or through the ESO archives. The data analysis code was written in \textsc{python} and may be shared upon reasonable written request to the corresponding author.



\bibliographystyle{mnras}
\bibliography{vandels_civ.bib} 








\bsp	
\label{lastpage}
\end{document}